\documentclass[orivec]{llncs}
\pdfoutput=1 

\usepackage[fleqn]{amsmath}
\usepackage{xfrac}
\usepackage{microtype}
\usepackage{xspace}

\usepackage[usenames,dvipsnames]{color}
\usepackage{algorithm2e}
\usepackage{silver}
\usepackage{suffix}
\usepackage[T1]{fontenc}
\usepackage{lmodern}
\usepackage{textcomp}
\IfFileExists{luximono.sty}{\usepackage[scaled=0.9]{luximono}}{\usepackage[scaled=0.81]{beramono}}
\usepackage{afterpage}
\usepackage{dcolumn}
\usepackage{makecell}
\usepackage{amssymb}

\usepackage{dcolumn}
\usepackage{makecell}

\def\istr{} 
\newcommand{\artifacturl}{https://doi.org/10.6084/m9.figshare.5900233}

\usepackage[firstpage]{draftwatermark}
\SetWatermarkText{\hspace*{4.5in}\raisebox{6.3in}{\includegraphics[scale=0.1]{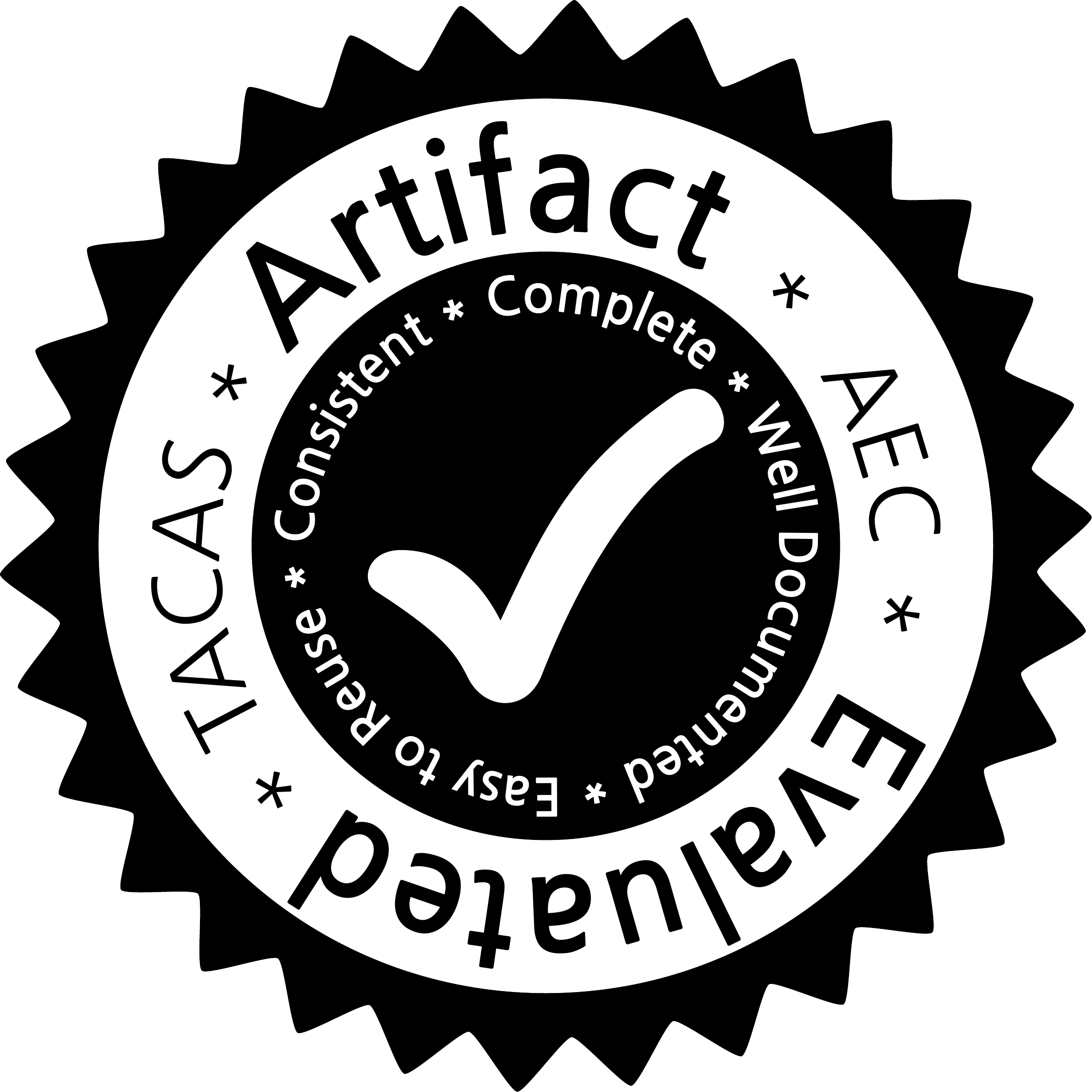}}}
\SetWatermarkAngle{0}

\usepackage[normalem]{ulem}

\newcommand{\blue}[1]{{\color{RoyalBlue}{#1}}}

\newcommand{\tick}{\ensuremath{\checkmark}}

\newcommand*{\figref}[1]{Fig.~\ref{fig:#1}}
\newcommand*{\secref}[1]{Sec.~\ref{sec:#1}}
\def\trref{\ifx\istr\undefined{ of the TR~\cite{TR}}\fi}
\def\trrefnocite{\ifx\istr\undefined{ of the TR}\fi}
\newcommand*{\appref}[1]{App.~\ref{sec:#1}\trref{}}
\newcommand*{\apprefnocite}[1]{App.~\ref{sec:#1}\trrefnocite{}}

\ifx\symlasy\undefined   \DeclareSymbolFont{lasy}{U}{lasy}{m}{n}
  \SetSymbolFont{lasy}{bold}{U}{lasy}{b}{n} \else
\fi
\DeclareMathSymbol\safeleadsto {\mathrel}{lasy}{"3B}

\newcommand*{\sil}[1]{\ifmmode\mbox{\inlinesilver{#1}}\else\inlinesilver{#1}\fi}

\newcommand*{\pointsto}[3]{\ensuremath{{#1}\overset{\scriptscriptstyle{#3}}{\mapsto}{#2}}}

\newcommand{\whole}[1]{\ensuremath{{\langle}{#1}{\rangle}}}
\newcommand{\conjuncts}[1]{\ensuremath{\cornerBrackets{#1}}}
\def\invariant{\operator{\textit{inv}}}
\def\inv{\invariant}

\def\down{\operator{\texttt{\downName}}}
\def\up{\operator{\texttt{\upName}}}
\def\temp{\operator{\texttt{\tempName}}}

\def\ceilBrackets#1{\ensuremath{\lceil#1\rceil}}

\def\slightindent{\;\;}

\def\tempName{tmp}
\def\downName{down}
\def\upName{up}
\def\realName{real}

\def\upAss#1{\ensuremath{\ceilBrackets{#1}^{\textit{\upName}}}}
\def\downAss#1{\ensuremath{\ceilBrackets{#1}^{\textit{\downName}}}}
\def\tempAss#1{\ensuremath{\ceilBrackets{#1}^{\textit{\tempName}}}}
\def\tempOrUpAss#1{\ensuremath{\ceilBrackets{#1}^{\textit{\tempName/\upName}}}}
\def\tempOrRealAss#1{\ensuremath{\ceilBrackets{#1}^{\textit{\tempName/\realName}}}}

\newcommand{\rlx}{\texttt{rlx}}
\newcommand{\na}{\texttt{na}}
\newcommand{\acq}{\texttt{acq}}
\newcommand{\rel}{\texttt{rel}}
\newcommand{\relacq}{\texttt{rel\_acq}}
\newcommand{\rmw}{\texttt{RMW}}

\newcommand{\alloc}{\texttt{alloc}}

\newcommand{\Assign}[2]{\ensuremath{#1\gets#2}}
\newcommand{\NonAtomicAlloc}[1]{\Assign{#1}{\alloc_{\na}()}}
\newcommand{\GhostAlloc}[1]{\Assign{#1}{\alloc_{\textit{ghost}}()}}
\newcommand{\AtomicAlloc}[3]{\Assign{#2}{\alloc_{#1}(#3)}}
\newcommand{\AtomicAllocAcq}[2]{\AtomicAlloc{\acq}{#1}{#2}}
\newcommand{\AtomicAllocRMW}[2]{\AtomicAlloc{\rmw}{#1}{#2}}

\newcommand{\Read}[3]{\Assign{#2}{[#3]_{#1}}}
\newcommand{\NonAtomicRead}[2]{\Read{\na}{#1}{#2}}
\newcommand{\AcquireRead}[2]{\Read{\acq}{#1}{#2}}
\newcommand{\RelaxedRead}[2]{\Read{\rlx}{#1}{#2}}

\newcommand{\Write}[3]{\Assign{[#2]_{#1}}{#3}}
\newcommand{\NonAtomicWrite}[2]{\Write{\na}{#1}{#2}}
\newcommand{\ReleaseWrite}[2]{\Write{\rel}{#1}{#2}}
\newcommand{\RelaxedWrite}[2]{\Write{\rlx}{#1}{#2}}

\newcommand{\Fence}{\texttt{fence}}

\newcommand{\Rewrite}[2]{\texttt{rewrite }#1\texttt{ as }#2}

\newcommand{\FenceAcq}{\ensuremath{\Fence_{\acq}}}
\def\FenceRel{\operator{\Fence_{\rel}}}

\def\while{\operator{\texttt{while}}}

\makeatletter
\def\CAS#1#2#3#4{\@ifnextchar\bgroup {\CASAssign{#1}{#2}{#3}{#4}}{\CASOp{#1}{#2}{#3}{#4}}}
\newcommand{\CASOp}[4]{\texttt{CAS}_{#1}(#2,#3,#4)}
\newcommand{\CASAssign}[5]{\Assign{#1}{\CASOp{#2}{#3}{#4}{#5}}}
\newcommand{\FetchAdd}[4]{\Assign{#1}{\texttt{fetch\_and\_add}_{#2}(#3,#4)}}
\newcommand{\FetchAddNo}[3]{\texttt{fetch\_and\_add}_{#1}(#2,#3)}

\newcommand{\triple}[3]{\ensuremath{\{{#1}\}\;{#2}\;\{{#3}\}}}
\newcommand{\doubleRowTriple}[5]{\ensuremath{\left\{{\begin{array}{c}#1\\#2\end{array}}\right\}\;{#3}\;\left\{{\begin{array}{c}#4\\#5\end{array}}\right\}}}
\newcommand{\emp}{\ensuremath{\textsf{emp}}}
\newcommand{\truesym}{\ensuremath{\textsf{true}}}
\newcommand{\falsesym}{\ensuremath{\textsf{false}}}
\newcommand{\Uninit}[1]{\ensuremath{\textsf{Uninit}(#1)}}
\newcommand{\Init}[1]{\ensuremath{\textsf{Init}(#1)}}
\newcommand{\Rel}[1]{\ensuremath{\textsf{Rel}(#1)}}
\newcommand{\Acq}[1]{\ensuremath{\textsf{Acq}(#1)}}
\newcommand{\RMWAcq}[1]{\ensuremath{\textsf{RMWAcq}(#1)}}
\newcommand{\InitX}{\ensuremath{\textsf{Init}}}
\newcommand{\RelX}{\ensuremath{\textsf{Rel}}}
\newcommand{\AcqX}{\ensuremath{\textsf{Acq}}}
\newcommand{\RMWAcqX}{\ensuremath{\textsf{RMWAcq}}}

\newcommand{\Q}{\ensuremath{\mathcal{Q}}}
\newcommand{\vvar}{\ensuremath{\mathcal{V}}}
\newcommand{\instantiate}[2]{\sub{#1}{#2}{\vvar}}
\newcommand{\upd}[2]{\ensuremath{(\vvar \neq #2 \Rightarrow #1)}}
\newcommand{\vecupd}[3]{\ensuremath{((\bigwedge_{#3}\vvar \neq #2) \Rightarrow #1)}}

\newcommand{\FSLpp}{\text{FSL\ensuremath{{+}{+}}}}

\newcommand{\entails}{\ensuremath{\models}}

\def\alt{\ \mid\ }

\newcommand{\cond}[3]{\ensuremath{(#1~{?}~#2~{:}~#3)}}

\def\UP{\spoperator{\bigtriangleup}{}}
\def\DOWN{\spoperator{\raisebox{2pt}{\ensuremath{\bigtriangledown}}}{}}

\newcommand{\numbers}{\textit{indices}}
\newcommand{\foreachSymb}{\textit{foreach}}
\newcommand{\foreach}[2]{\textit{\foreachSymb~{#1}~in~{#2}}}
\newcommand{\foreachNumber}{\foreach{i}{\numbers}}
\newcommand{\foreachStart}{\textit{do}}
\newcommand{\foreachEnd}{\textit{end}}

\newcommand{\env}{\sigma}

\def\heap{\operator{\texttt{heap}}}

\def\myvspace{\vspace}
\newcommand{\eg}{{{e.g.\@}}}
\newcommand{\ie}{{{i.e.\@}}}
\newcommand{\cf}{{\it{cf.\@}}}

\newcommand{\notapplicable}{\text{n/a}}

\renewcommand*{\gets}{~{:=}~}%

\newcommand*{\FV}{\operator{\ensuremath{\mathit{FV}}}}

\newcommand{\sub}[3]{\ensuremath{#1[#2/#3]}}

\newcommand{\hlLine}[1]{\leavevmode\rlap{\hbox to \hsize{\color{Lightgray}\leaders\hrule height .8\baselineskip depth .9ex\hfill}}#1}
\newsavebox{\newBox}
\newlength{\newWidth}
\newcommand{\hl}[1]{
\savebox{\newBox}{\ensuremath{#1}}\settowidth{\newWidth}{\usebox{\newBox}}\leavevmode\rlap{\hbox to \hsize{\color{Lightgray}\leaders\hrule height .8\baselineskip depth .9ex\hskip \newWidth}}\usebox{\newBox}}

\newcommand{\safecode}[1]{\text{\code{#1}}}

\usepackage{prooftree}

\makeatletter

\newskip \point

\def \premisespacing{\quad}

\def \RulePremisesNewlineMore[#1]#2.#3#4{\@ifnextchar\bgroup{\RulePremisesNewlineMore[#1]{#2}.{#3\premisespacing#4}}{\@ifnextchar.{\RulePremisesNewline[#1]{{\begin{array}{c}#2\\#3\premisespacing#4\end{array}}}}{\RuleMultiPremise[#1]{{\begin{array}{c}#2\\#3\end{array}}}{#4}}}}

\def \RulePremisesNewline[#1]#2.#3{\@ifnextchar\bgroup{\RulePremisesNewlineMore[#1]{#2}.{#3}}{\@ifnextchar.{\RulePremisesNewline[#1]{{\begin{array}{c}#2\\#3\end{array}}}}{\RuleMultiPremise[#1]{#2}{#3}}}}

\def \RuleMultiPremise[#1]#2#3{\@ifnextchar\bgroup{\RuleMultiPremise[#1]{#2\premisespacing#3}}{\@ifnextchar.{\RulePremisesNewline[#1]{#2\premisespacing#3}}{\prooftree #2\justifies#3 \using{#1}\endprooftree}}}

\def \RuleWithName[#1]#2{\@ifnextchar\bgroup {\RuleMultiPremise[#1]{#2}}{\@ifnextchar.{\RulePremisesNewline[#1]{#2}}{\prooftree \justifies #2 \using{#1} \endprooftree}}}

\def \RuleWithInfo[#1]{\@ifnextchar[{\RuleWithNameAndCondition[#1]}{\RuleWithName[(#1)]}}

\def \RuleWithNameAndCondition[#1][#2]{\RuleWithName[(#1)^{#2}]}

\def \Inf{\proofrulebaseline=2ex \abovedisplayskip12\point\belowdisplayskip12\point \abovedisplayshortskip8\point\belowdisplayshortskip8\point \@ifnextchar[{\RuleWithInfo}{\RuleWithName[ ]}}

\makeatother

\makeatletter
\def\operator#1{\@ifnextchar\bgroup {\operatorarg{\ensuremath{#1}}}{\ensuremath{#1}}}
\def\operatorarg#1#2{{#1}{\ensuremath{(#2)}}}
\def\spoperator#1#2{\@ifnextchar\bgroup{\spoperatorarg{\ensuremath{#1}}{\ensuremath{#2}}}{\ensuremath{#1}}}
\def\spoperatorarg#1#2#3{\ensuremath{#1#2#3}}

\def\fixedoperator#1{\@ifnextchar\bgroup {\fixedoperatorarg{#1}}{\ensuremath{#1}}}
\def\fixedoperatorarg#1#2{\fixedoperatorparse{#1}#2~}
\def\fixedoperatorparse#1#2,#3~{\ensuremath{{#2}{.}{#1}{(#3)}}}
\makeatother

\newskip \point \point =1pt
\setbox134=\hbox{\leavevmode\raise0\point\hbox{${\langle}\kern-2.5\point{\langle}$}}
\setbox135=\hbox{\raise0\point\hbox{${ \rangle}\kern-2.5\point{ \rangle}$}}
\def \cornerBrackets#1{\copy134{#1}\copy135}
\setbox136=\hbox{\leavevmode\raise0\point\hbox{${\lfloor}\kern-3.0\point{\lfloor}$}}
\setbox137=\hbox{\raise0\point\hbox{${ \rfloor}\kern-3.0\point{ \rfloor}$}}
\newcommand{\floorBrackets}[2]{\copy136{#1}\copy137_{#2}}
\setbox138=\hbox{\leavevmode\raise0\point\hbox{${[}\kern-1.5\point{[}$}}
\setbox139=\hbox{\raise0\point\hbox{${]}\kern-1.5\point{]}$}}
\newcommand{\sqBrackets}[2]{\copy138{#1}\copy139_{#2}}

\def \encodeAss#1{\floorBrackets{#1}{}}
\def \encodeStm#1{\sqBrackets{#1}{}}
\def \concrete#1#2#3#4{\cornerBrackets{#1}_{#2,#3,#4}}
\newcommand{\eval}[4]{\ensuremath{\langle{#1}\rangle_{#2,#3,#4}}} 
\newcommand{\comment}[1]{}
\begin{document}

\title{Automating Deductive Verification\\for Weak-Memory Programs\ifdefined\istr\\(extended version)\fi}

\author{Alexander J. Summers \and Peter M\"uller}

\institute{Department of Computer Science, ETH Zurich, Switzerland \\
\email{\{alexander.summers, peter.mueller\}@inf.ethz.ch}
}
\maketitle
\pagestyle{plain}

\begin{abstract}
Writing correct programs for weak memory models such as the C11 memory
model is challenging because of the weak consistency guarantees these
models provide. The first program logics for the verification of such
programs have recently been proposed, but their usage has been limited
thus far to manual proofs.  Automating proofs in these logics via
first-order solvers is non-trivial, due to features such as
higher-order assertions, modalities and rich permission resources.

In this paper, we provide the first encoding of a weak memory
program logic using existing deductive verification tools. Our work
enables, for the first time, the (unbounded) verification of C11 programs
at the level of abstraction provided by the program logics; the only
necessary user interaction is in the form of specifications written in
the program logic.

We tackle
three recent program logics: Relaxed Separation Logic and two forms of Fenced
Separation Logic, and show how these can be encoded using the Viper
verification infrastructure. In doing so, we illustrate several novel
encoding techniques which could be employed for other logics. Our work is
implemented, and has been evaluated on examples from existing papers as well
as the Facebook open-source Folly library.
\end{abstract}

%
%
%
%
%
%

\section{Introduction}
\label{sec:introduction}

Reasoning about programs running on weak memory is challenging because
weak memory models admit executions that are not sequentially
consistent, that is, cannot be explained by a sequential interleaving
of concurrent threads. Moreover, weak-memory programs employ a range
of operations to access memory, which require dedicated reasoning
techniques.  These operations include fences as well as read and write
accesses with varying degrees of synchronisation.

Some of these challenges are addressed by the first program logics for
weak-memory programs, in particular, Relaxed Separation Logic
(RSL)~\cite{VafeiadisN13}, GPS~\cite{TuronVD14}, Fenced Separation Logic
(FSL)~\cite{DokoV16}, and \FSLpp{}~\cite{DokoVafeiadis17}. These logics
apply to interesting classes of C11 programs, but their tool support
has been limited to embeddings in Coq. Verification based on these
embeddings requires substantial user interaction, which is
an obstacle to applying and evaluating these logics.

\begin{figure}[t]
\begin{center}
\scalebox{0.9}{
\[
\begin{array}{rcl}
s\quad&{::=}&\NonAtomicAlloc{l}\alt\AtomicAlloc{\rho}{l}{\Q}\alt\Write{\sigma}{l}{e}\alt\Read{\sigma}{x}{l}\\
&\alt&\FenceAcq\alt\FenceRel{A}\alt\CAS{x}{\tau}{l}{e_1}{e_2}\\
&&\textit{where }\rho\in\{\acq,\rmw\}\textit{,}\quad\sigma\ {::=}\ \na\mid\tau\textit{,}\quad\tau\in\{\acq,\rel,\relacq,\rlx\}
\end{array}
\]
}
\end{center}
\myvspace{-3mm}
\caption{Syntax for memory accesses. \na{} indicates a non-atomic operation; $\tau$ indicates an atomic access mode (as defined in C11), discussed in later sections. $\rho$, and assertions $A$ and invariants $\Q$ are program annotations, needed as input for our encoding.
Expressions $e$ include boolean and arithmetic operations,
but no heap accesses. We assume that source programs are type-checked.}
\label{fig:syntaxone}
\end{figure}

In this paper, we present a novel approach to automating deductive
verification for weak memory programs. We encode large fractions of
RSL, FSL, and \FSLpp{} (collectively referred to
as \emph{the RSL logics}) into the intermediate verification language
Viper~\cite{MuellerSchwerhoffSummers16}, and use the existing Viper
verification backends to reason automatically about the encoded
programs. This encoding reduces all concurrency and weak-memory
features as well as logical features such as higher-order assertions
and custom modalities to a much simpler sequential logic.

Defining an encoding into Viper is much more lightweight than
developing a dedicated verifier from scratch, since we can reuse the
existing automation for a variety of advanced program reasoning
features.  Compared to an embedding into an interactive theorem prover
such as Coq, our approach leads to a significantly higher degree of
automation than that typically achieved through tactics. Moreover, it
allows users to interact with the verifier on the abstraction level of
source code and annotations, without exposing the underlying
formalism.  Verification in Coq can provide foundational guarantees,
whereas in our approach, errors in the encoding or bugs in the
verifier could potentially invalidate verification results. We
mitigate the former risk by a soundness argument for our encoding and
the latter by the use of a mature verification system. We are
convinced that both approaches are necessary: foundational
verification is ideal for meta-theory development and application
areas such as safety-critical systems, whereas our approach is
well-suited for prototyping and evaluating logics, and for making a
verification technique applicable by a wider user base.

The contributions of this paper are:
(1)~The first automated deductive verification approach for weak-memory logics.
We demonstrate the effectiveness of this approach on examples from
the literature, which are available online~\cite{OnlineAppendix}.
(2)~An encoding of large fractions of RSL, FSL, and \FSLpp{} into Viper. Various
aspects of this encoding (such as the treatment of higher-order
features and modalities, as well as the overall proof search strategy)
are generic and can be reused to encode other advanced separation logics.
(3)~A prototype implementation, which is available online \cite{PrototypeTool}.
%

\subsubsection{Related Work.}
The existing weak-memory logics RSL~\cite{VafeiadisN13},
GPS~\cite{TuronVD14}, FSL~\cite{DokoV16}, and \FSLpp{}~\cite{DokoVafeiadis17} have
been formalized in Coq and used to verify small examples.
The proofs were constructed mostly manually, whereas our approach
automates most of the proof steps. As shown in
 our evaluation, our approach reduces the overhead by
more than an order of magnitude. The degree of automation in Coq
could be increased through logic-specific tactics~(\eg{} \cite{Sergey15,Chlipala11}),
whereas our approach benefits from Viper's automation for the intermediate
language, which is independent of the encoded
logic.

Jacobs~\cite{KULeuven-452373} proposed a program logic for the TSO
memory model that has been encoded in VeriFast~\cite{Jacobs-VeriFast10}. This encoding requires a substantial amount of annotations, whereas
our approach provides a higher degree of automation and handles the more
complex C11 memory model.

Weak-memory reasoning has been addressed using techniques based on
model-checking (\eg{} \cite{Bouajjani13,Abdulla16,Abdulla17}) and static analyses (\eg{}
\cite{Dan15,Alglave17}). These approaches are fully automatic, but do not
analyse code modularly, which is \eg{} important for verifying
libraries independently from their clients. Deductive verification enables compositional proofs by
requiring specifications at function boundaries. Such specifications can preserve arbitrarily-precise information about the (unbounded) behaviour of a program's constituent parts.

Automating logics via encodings into intermediate verification
languages is a proven approach, as witnessed by
the many existing verifiers
(\eg{} \cite{Cohen09,Cuoq12,Leino:2010:DAP:1939141.1939161,LeinoMueller09})
which target Boogie~\cite{Barnett05} or Why3~\cite{boogie11why3}. Our
work is the first that applies this approach to logics for weak-memory
concurrency. Our encoding benefits from Viper's native support
for separation-logic-style reasoning
and several advanced features such as quantified permissions
and permission introspection~\cite{MuellerSchwerhoffSummers16,MuellerSchwerhoffSummers16b}, which are not available in other intermediate
verification languages.

\subsubsection{Outline.}
The next four sections present our encoding for the core features
of the C11 memory model: we discuss non-atomic locations in
\secref{non-atomics}, release-acquire accesses in
\secref{release-acquire}, fences in \secref{fences}, and
compare-and-swap in \secref{compare-and-swap}. We discuss
soundness and completeness of our encoding
in \secref{soundness} and
evaluate our approach in \secref{examples}.
\secref{conclusion} concludes. Further details of our encoding and examples are available in \ifdefined\istr the appendix. \else our accompanying technical report (hereafter, TR) \cite{TR}. \fi
A prototype implementation of our encoding (with all examples) is available as an artifact~\cite{PrototypeTool}.

\begin{figure}[t]
\begin{center}
\scalebox{0.9}{
\[
\begin{array}{rcl}
A \quad &{::=}& e \alt \pointsto{l}{e}{k} \alt A_1 \ast A_2 \alt e\Rightarrow A \alt \cond{e}{A_1}{A_2}\\ 
&\alt&\Uninit{l}\alt\Acq{l,\Q}\alt\Rel{l,\Q}\alt\Init{l}\alt\UP{A}\alt\DOWN{A}\alt\RMWAcq{l,\Q}
\end{array}
\]
}
\end{center}
\myvspace{-3mm}
\caption{Assertion syntax of the RSL logics. The top row of constructs are standard for separation logics; those in the second row are specific to the RSL logics, and explained throughout the paper. Invariants \Q{} are \emph{functions} from values to assertions (cf.~\secref{release-acquire}).}
\label{fig:assertions-RSL}
\end{figure}

\section{Non-atomic Locations}
\label{sec:non-atomics}

We present our encoding for a small imperative programming language
similar to the languages supported by the RSL
logics.  C11 supports \emph{non-atomic} memory accesses and different
forms of \emph{atomic} accesses. The access operations are summarised
in \figref{syntaxone}.  We adopt the common simplifying
assumption~\cite{VafeiadisN13,TuronVD14} that memory locations are
partitioned into those accessed only via non-atomic accesses
(\emph{non-atomic locations}), and those accessed only via C11 atomics
(\emph{atomic locations}).  Read and write statements are
parameterised by a mode $\sigma$, which is either \na{} (non-atomic)
or one of the atomic access modes 
   $\tau$.
We focus on non-atomic accesses in this section and discuss atomics in
subsequent sections.

\subsubsection{RSL proof rules.}

\begin{figure}[t]
\begin{center}
\scalebox{0.9}{
\[
\begin{array}{cc}
\Inf{\vdash\triple{\truesym}{\NonAtomicAlloc{l}}{\Uninit{l}}}\quad\quad&\quad\quad
\Inf{\vdash\triple{\pointsto{l}{\_}{1}\vee\Uninit{l}}{\NonAtomicWrite{l}{e}}{\pointsto{l}{e}{1}}}\\[3mm]
\Inf{\vdash\triple{\pointsto{l}{e}{k}}{\NonAtomicRead{x}{l}}{x=e \ast \pointsto{l}{e}{k}}}\quad\quad&\quad\quad
\vspace{4pt}(\pointsto{l}{e}{k}\ \ast\ \pointsto{l}{e'}{k'}) \Leftrightarrow (e = e' \ast \pointsto{l}{e}{k + k'})
\end{array}
\]
}
\end{center}
\myvspace{-3mm}
\caption{Adapted RSL rules for non-atomics. Read access requires a non-zero permission.
Write access requires either write permission or that the location is
uninitialised. The underscore $\_$ stands for an arbitrary value.}
\label{fig:nonatomic-RSL}
\end{figure}
Non-atomic memory accesses come with no synchronisation guarantees;
programmers need to ensure that all accesses to non-atomic
locations are data-race free. The RSL logics enforce this requirement
using standard separation logic~\cite{OHearn01,Reynolds02a}. We show
the syntax of assertions in \figref{assertions-RSL}, which will be
explained throughout the paper. A \emph{points-to assertion}
$\pointsto{l}{e}{k}$ denotes a transferrable
\emph{resource}, providing permission to access the location $l$, and
expressing that $l$ has been initialised and its current
value is $e$. Here, $k$ is a fraction $0 < k \leq 1$;
$k=1$ denotes the \emph{full} (or exclusive)
permission to read and write location $l$, whereas $0 < k < 1$
provides (non-exclusive) read access~\cite{Boyland}. Points-to resources can be split
and recombined, but never duplicated or forged; when
transferring such a resource to another thread it is
removed from the current one, avoiding data races by
construction. The RSL assertion
$\Uninit{l}$ expresses exclusive access to a location $l$ that
has been allocated, but not yet initialised; $l$ may be
written to but not read from.
 The main proof rules for non-atomic locations, adapted from RSL~\cite{VafeiadisN13}, are shown in \figref{nonatomic-RSL}.

\subsubsection{Encoding.}
The Viper intermediate verification language
\cite{MuellerSchwerhoffSummers16}
supports an assertion language based on Implicit Dynamic
Frames~\cite{Smans:2012}, a program logic related to separation
logic~\cite{ParkinsonSummers12}, but which separates permissions from
value information. Viper is object-based; the only memory locations
are field locations $e.f$ (in which $e$ is a reference, and $f$ a
field name). Permissions to access these heap locations are described
by \emph{accessibility predicates} of the form \code{acc($e$.$f$, $k$)},
where $k$ is a fraction as for points-to predicates
above ($k$ defaults to $1$). Assertions that do not contain accessibility predicates are
called \emph{pure}. Unlike in separation logics, heap locations may
be read in pure assertions.

We model C-like memory locations $l$ using a field \code{val} of a
Viper reference $l$. Consequently, a separation logic assertion
$\pointsto{l}{e}{k}$ is represented in Viper as
\code{acc($l$.val, $k$) && $l$.val == e}. We assume
that memory locations have type \code{int}, but a generalisation is
trivial.
Viper's conjunction \code{&&} treats
permissions like a separating conjunction, requiring the sum of the permissions in
each conjunct, and acts as logical conjunction for pure assertions
(just as $\ast$ in separation logic).

Viper provides two key statements for encoding proof rules:
\code{inhale $A$}
\emph{adds} the permissions denoted by the assertion $A$ to
the current state, and \emph{assumes} pure assertions in $A$. This
can be used to model gaining new resources, \eg{}, acquiring a lock in the source program. Dually, \code{exhale $A$} checks that the current
state satisfies $A$ (otherwise a verification error occurs), and
\emph{removes} the permissions that $A$ denotes; the values of any
locations to which no permission remains are \emph{havoced}
(assigned arbitrary values). For example, when forking a new thread,
its precondition is exhaled to transfer
the necessary resources from the forking thread.  Inhale
and exhale statements can be seen as the permission-aware analogues of
the assume and assert statements of first-order verification languages~\cite{LeinoMueller09}.

\begin{figure}[t]
\begin{silver}[mathescape=true]
field val:  Int
field init: Bool

$\encodeAss{\Uninit{l}}\leadsto\ $acc(l.val) && acc(l.init) && !l.init
$\encodeAss{\pointsto{l}{e}{k}}\leadsto\ $acc(l.val, k) && acc(l.init, k) && l.val == $\encodeAss{e}$ && l.init

$\encodeStm{\NonAtomicAlloc{l}}\leadsto\ $l := new(); inhale $\encodeAss{\Uninit{l}}$
$\encodeStm{\NonAtomicRead{x}{l}}\leadsto\ $assert l.init; x := l.val
$\encodeStm{\NonAtomicWrite{l}{e}}\leadsto\ $l.val  := $\encodeAss{e}$; l.init := true
\end{silver}
\myvspace{-0.75em}
\caption{Viper encoding of the RSL assertions and the rules for non-atomic memory accesses from \figref{nonatomic-RSL}. 
}
\label{fig:nonatomic-Viper}
\end{figure}

The encoding of the rules for non-atomics from \figref{nonatomic-RSL}
is presented in \figref{nonatomic-Viper}. $\encodeAss{A}\leadsto \ldots$
denotes the encoding of an RSL assertion $A$ as a
Viper assertion, and analogously $\encodeStm{s} \leadsto \ldots$ for
source-level statements $s$.

The first two lines show background declarations.
The assertion encodings follow the explanations
above. Allocation is modelled by obtaining a fresh reference (via
\code{new()}) and inhaling permissions to its \code{val} and
\code{init} fields; assuming \code{!l.init} reflects that the location
is not yet initialised. Viper implicitly checks the necessary
permissions for field accesses (verification fails otherwise). Hence,
the translation of a non-atomic read only needs to check that the
read location is initialised before obtaining its value. Analogously,
the translation of a non-atomic write only stores the value and records
that the location is now initialised.

Note that Viper's implicit permission checks are both necessary and
sufficient to encode the RSL rules in \figref{nonatomic-RSL}. In
particular, the assertions $\pointsto{l}{\_}{1}$ and $\Uninit{l}$ both
provide the permissions to write to location $l$. By including
\code{acc(}$l$.\code{val)} in the encoding of both assertions, we
avoid the disjunction of the RSL rule.

Like the RSL logics, our approach requires programmers to annotate their code with access
modes for locations (as part of the \code{alloc} statement), and specifications such as pre and postconditions for methods and threads. Given these inputs, Viper constructs the proof
automatically. In particular, it automatically proves entailments,
and splits and combines fractional permissions (hence, the equivalence
in \figref{nonatomic-RSL} need not be encoded). Automation can be increased
further by inferring some of the required assertions, but this is
orthogonal to the encoding presented in this paper.

\section{Release-Acquire Atomics}
\label{sec:release-acquire}

\begin{figure}[t]
\begin{center}
\scalebox{0.9}{
\[
\begin{array}{c}
\blue{\Q_1\ \equiv\ (\vvar \neq 0 \Rightarrow \pointsto{a}{42}{1})\quad\quad \Q_2\ \equiv\ (\vvar \neq 0 \Rightarrow \pointsto{b}{7}{1})}\\[5pt]
\blue{\{\truesym\}}\\ 
\NonAtomicAlloc{a};\ \NonAtomicAlloc{b};\ \AtomicAllocAcq{l}{\blue{\Q_1{\ast}\Q_2}};\ \ReleaseWrite{l}{0}\\
\begin{array}{l}
\blue{\{\Acq{l,\Q_1}\ast\Init{l}\}}\\
\slightindent\while{[l]_{\acq} == 0};\\
\slightindent\NonAtomicRead{x}{a}\\
\slightindent\NonAtomicWrite{a}{x + 1}\\
\blue{\{\truesym\ast \pointsto{a}{43}{1}\}}
\end{array}
\left|
\left|\
\begin{array}{l}
\blue{\{\Uninit{a}\ast\Uninit{b}\ast\Rel{l,\Q_1{\ast}\Q_2}\}}\\
\slightindent\NonAtomicWrite{a}{42}\\
\slightindent\NonAtomicWrite{b}{7}\\
\slightindent\ReleaseWrite{l}{1}\\
\blue{\{\truesym\ast\Init{l}\}}
\end{array}
\right|
\right|\
\begin{array}{l}
\blue{\{\Acq{l,\Q_2}\ast\Init{l}\}}\\
\slightindent\while{[l]_{\acq} == 0};\\
\slightindent\NonAtomicRead{y}{b}\\
\slightindent\NonAtomicWrite{b}{y + 1}\\
\blue{\{\truesym\ast \pointsto{b}{8}{1}\}}
\end{array}\\
\blue{\{\truesym \ast \pointsto{a}{43}{1} \ast \pointsto{b}{8}{1} \ast \Init{l}\}}
\end{array}
\]}
\end{center}
\myvspace{-4mm}
\caption{An example illustrating ``message passing'' of non-atomic ownership, using release acquire
atomics (inspired by an example from \cite{DokoV16}). Annotations are shown in \blue{blue}. This example corresponds to \code{RelAcqDblMsgPassSplit} in our evaluation (\secref{examples}).}
\label{fig:release-acquire-example-RSL}
\end{figure}

The simplest form of C11 atomic memory accesses are \emph{release
  write} and \emph{acquire read} operations.  They can be used to
synchronise the transfer of ownership of (and information about)
other, non-atomic locations, using a \emph{message passing idiom},
illustrated by the example in
\figref{release-acquire-example-RSL}. This program allocates two
non-atomic locations $a$ and $b$, and an atomic location $l$
(initialised to $0$), which is used to synchronise the three threads
that are spawned afterwards. The middle thread makes changes to the
non-atomics $a$ and $b$, and then signals completion via a release
write of $1$ to $l$; conceptually, it gives up ownership of the
non-atomic locations via this signal. The other threads loop
attempting to acquire-read a non-zero value from $l$. Once they do,
they each gain ownership of one non-atomic location via
the acquire read of $1$ and access that location.
The release write and acquire reads of value $1$ enforce
\emph{ordering constraints} on the non-atomic accesses,
preventing the left and right threads from racing with the middle
one.

\subsubsection{RSL proof rules.}
The RSL logics capture message-passing idioms by associating a
\emph{location invariant} $\Q$ with each atomic location. Such an
invariant is a function from values to assertions; we represent such
functions as assertions with a distinguished variable symbol $\vvar$
as parameter. Location invariants prescribe the intended ownership
that a thread obtains when performing an acquire read of value $\vvar$
from the location, and that must correspondingly be given up by a
thread performing a release write. The main proof
rules~\cite{VafeiadisN13} are shown in \figref{release-acquire-RSL}.

When allocating an atomic location for release/acquire accesses (first
proof rule), a location invariant $\Q$ must be chosen (as an
annotation on the allocation). The assertions $\Rel{l,\Q}$ and
$\Acq{l,\Q}$ record the invariant to be used with subsequent
release writes and acquire reads. To perform a release write of value
$e$ (second rule), a thread must hold the $\Rel{l,\Q}$ assertion and
\emph{give up} the assertion $\instantiate{\Q}{e}$. For example, the
line $\ReleaseWrite{l}{1}$ in \figref{release-acquire-example-RSL}
causes the middle thread to give up ownership of both non-atomic
locations $a$ and $b$. The assertion $\Init{l}$ represents that atomic
location $l$ is initialised; both $\Init{l}$ and $\Rel{l,\Q}$ are
\emph{duplicable} assertions; once obtained, they can be passed to
multiple threads.

Multiple acquire reads might read the value written by a single
release write operation; RSL prevents ownership of the transferred
resources from being obtained (unsoundly) by multiple readers
in two ways. First,
$\Acq{l,\Q}$ assertions cannot be
duplicated, only split by partitioning the invariant $\Q$ into
disjoint parts. For example, in
\figref{release-acquire-example-RSL}, $\Acq{l,\Q_1}$ is given to the
left thread, and $\Acq{l,\Q_2}$ to the right. Second,
the rule for acquire reads adjusts the invariant
in the $\AcqX$ assertion such that
subsequent reads of the same value will not obtain any ownership.

\begin{figure}[t]
\myvspace{-4mm}
\begin{center}
\scalebox{0.9}{
\[
\begin{array}{c}
\Inf{\vdash\triple{\truesym}{\AtomicAllocAcq{l}{\Q}}{\Rel{l,\Q} \ast \Acq{l,\Q}}}\\[3mm]
\Inf{\vdash\triple{\Q(e) \ast \Rel{l,\Q}}{\ReleaseWrite{l}{e}}{\Init{l}\ast \Rel{l,\Q}}}\\[3mm]
\Inf{\vdash\triple{\Init{l} \ast \Acq{l,\Q}}{\AcquireRead{x}{l}}{\instantiate{\Q}{x} \ast \Acq{l,\upd{\Q}{x}}}}\\[4mm]
\Init{l}  \ \Leftrightarrow\   \Init{l} \ast \Init{l}\quad\quad
\Rel{l,\Q}  \ \Leftrightarrow\  \Rel{l,\Q} \ast \Rel{l,\Q}\\[3mm]
\Acq{l,\Q_1 \ast \Q_2}  \ \Leftrightarrow\   \Acq{l,\Q_1} \ast \Acq{l,\Q_2}\quad\quad
\Q_1 \entails \Q_2 \ \Rightarrow\ \Acq{l,\Q_1} \entails \Acq{l,\Q_2}
\end{array}
\]
}
\end{center}
\myvspace{-3mm}
\caption{Adapted RSL rules for release-acquire atomics.
}
\label{fig:release-acquire-RSL}
\end{figure}

\subsubsection{Encoding.}
A key challenge for encoding the above proof rules is that $\RelX$
and $\AcqX$ are parameterised by the invariant $\Q$;
higher-order assertions are not directly supported in Viper. However,
for a given program, only finitely many such
parameterisations will be required, which allows us to apply
defunctionalisation~\cite{reynolds1972}, as follows. Given an
annotated  program,
we assign a unique \emph{index} to
each syntactically-occurring invariant $\Q$ (in particular, in
allocation statements, and as parameters to $\RelX$ and $\AcqX$
assertions in specifications). Furthermore, we assign unique indices to all \emph{immediate
  conjuncts} of these invariants.
We write $\numbers$ for the set of
indices used. For each $i$ in $\numbers$, we write $\inv{i}$ for the
invariant which $i$ indexes. For an invariant $\Q$,
we write $\whole{\Q}$ for its index, and $\conjuncts{\Q}$ for
the set of indices assigned to its immediate conjuncts.

Our encoding of the RSL rules from \figref{release-acquire-RSL} is
summarised in \figref{release-acquire-Viper}.  To encode duplicable
assertions such as $\Init{l}$, we make use of Viper's \emph{wildcard
  permissions}~\cite{MuellerSchwerhoffSummers16}, which represent
unknown positive permission amounts. When exhaled, these amounts are
chosen such that the amount exhaled will
be \emph{strictly smaller} than the amount held (verification fails
if no permission is held)~\cite{HeuleLeinoMuellerSummers13}.
So after inhaling an
$\Init{l}$ assertion (that is, a \code{wildcard} permission), it is possible
to exhale two \code{wildcard} permissions, corresponding to two
$\Init{l}$ assertions. Note that for atomic locations, we only use the
\code{init} field's permissions, not its value.

We represent a $\Rel{l,\_}$ assertion for \emph{some}
invariant via a \code{wildcard} permission to a \code{rel} field; this
is represented via the \code{SomeRel(l)} macro\footnote{Viper macros
  can be defined for assertions or statements, and are syntactically
  expanded (and their arguments substituted) on use.}, and is used as
the precondition for a release write (we must hold \emph{some} $\RelX$ assertion, according to \figref{release-acquire-RSL}). The specific
invariant associated with the location $l$ is represented by storing
its index as the \emph{value} of the \code{rel} field; when
encoding a release write, we branch on this value to exhale the
appropriate assertion.

\begin{figure}[t]
\begin{silver}[mathescape=true]
field rel: Int  $
$
field acq: Bool $
$
predicate AcqConjunct(l: Ref, idx: Int)

function valsRead(l: Ref, i: Int): Set[Int]
  requires AcqConjunct(l, i)

define SomeRel(l)  acc(l.rel, wildcard)
define SomeAcq(l)  acc(l.acq, wildcard) && l.acq == true

$\encodeAss{\Init{l}} \leadsto$ acc(l.init, wildcard)
$\encodeAss{\Rel{l,\Q}} \leadsto$ SomeRel(l) && l.rel == $\whole{\Q}$
$\encodeAss{\Acq{l,\Q}} \leadsto$ SomeAcq(l) && ($\foreach{i}{\conjuncts{\Q}}$:
  AcqConjunct(l, $i$) && valsRead(l, $i$) == Set[Int]() $\foreachEnd$)

$\encodeStm{\AtomicAllocAcq{l}{\Q}} \leadsto
$ l := new(); inhale $\encodeAss{\Rel{l,\Q}}$ && $\encodeAss{\Acq{l,\Q}}$

$\encodeStm{\ReleaseWrite{l}{e}} \leadsto
$ assert SomeRel(l);
  $\foreachNumber$ $\foreachStart$
    if ($i$ == l.rel) { exhale $\instantiate{\invariant{i}}{e}$  }
  $\foreachEnd$
  inhale Init(l)

$\encodeStm{\AcquireRead{x}{l}} \leadsto
$ assert Init(l) && SomeAcq(l); x := havoc(); // unknown Int
  $\foreachNumber$ $\foreachStart$
    if (perm(AcqConjunct(l, $i$)) == 1 && !(x in valsRead(l, $i$))) {
      inhale $\instantiate{\invariant{i}}{\texttt{x}}$
      tmpSet := valsRead(l, $i$)
      exhale AcqConjunct(l, $i$)
      inhale AcqConjunct(l, $i$) && valsRead(l,$i$) == tmpSet union Set(x)
    }
  $\foreachEnd$

\end{silver}
\myvspace{-0.9em}
\caption{Viper encoding of the RSL rules for release-acquire atomics
from \figref{release-acquire-RSL}.
  The operations in italics (\eg{} \texttt{$\foreachSymb$}) are expanded statically in our encoding into
conjunctions or statement sequences.
 The  value of the \code{acq} field will be explained in \secref{compare-and-swap}.}
\label{fig:release-acquire-Viper}
\myvspace{-4mm}
\end{figure}

Analogously to $\RelX$, we represent an $\AcqX$ assertion for
\emph{some} invariant using a \code{wildcard} permission (the
\code{SomeAcq} macro), which is the precondition for executing an
acquire read. However, to support splitting, we represent the
invariant in a more fine-grained way, by recording individual
conjuncts separately.  Each conjunct $i$ of the invariant is modelled
as an abstract \emph{predicate} instance \code{AcqConjunct($l$, $i$)},
which can be inhaled and exhaled individually.
This encoding handles the common case that invariants
are split along top-level conjuncts, as in \figref{release-acquire-example-RSL}.
More complex splits can be supported through additional annotations: see
\appref{rewriting-invariants}.

A release write is encoded by checking that some $\RelX$
assertion is held, and then exhaling the associated invariant for the
value written. Moreover, it records that the location is initialised.
The RSL rule for acquire reads adjusts the $\AcqX$ invariant by
obliterating the assertion for the value read. Instead of directly
representing the adjusted invariant (which would complicate our numbering
scheme), we track the set of values read as
state in our encoding. We complement each \code{AcqConjunct} predicate instance
with an (uninterpreted) Viper function \code{valsRead($l$, $i$)},
returning a set of indices\footnote{Viper's heap-dependent functions
  are mathematical functions of their parameters and the resources
  stated in their preconditions (here, \texttt{AcqConjunct($l$,$i$)}).}.

An acquire read checks that the location is initialised and that we
have \emph{some} $\AcqX$ assertion for the location. It assigns an
unknown value to the lhs variable $x$, which is subsequently constrained
by the invariant associated with the $\AcqX$ assertion as follows:
We check for each index whether we both currently hold an
\code{AcqConjunct} predicate for that index\footnote{A \code{perm}
expression yields the permission fraction held for a field or predicate
instance.}, and if so,
have not previously read the value $x$ from that conjunct of
our invariant. If these checks succeed,
we inhale the indexed
invariant for  $x$, and then include $x$ in the values read.

The encoding presented so far allows us to automatically verify
annotated C11 programs using release writes and acquire reads (\eg{},
the program of \figref{release-acquire-example-RSL}) without any
custom proof strategies~\cite{OnlineAppendix}. In particular, we can
support the higher-order $\AcqX$ and $\RelX$ assertions through
defunctionalisation and enable the splitting of invariants through a
suitable representation.

%
%
%
%

\section{Relaxed Memory Accesses and Fences}
\label{sec:fences}

In contrast to release-acquire accesses, C11's \emph{relaxed} atomic
accesses provide no synchronisation: threads may
observe reorderings of relaxed accesses and other memory
operations.
Correspondingly, RSL's proof rules for relaxed atomics
provide weak guarantees, and do not support ownership
transfer.
%
Memory fence instructions can eliminate this problem. Intuitively, a
\emph{release fence} together with a subsequent relaxed write allows a
thread to transfer away ownership of resources, similarly to a release
write.
Dually, an \emph{acquire
  fence} together with a prior relaxed read allows a thread to obtain
ownership of resources, similarly to an acquire read.
This reasoning is justified by the ordering
guarantees of the C11 model~\cite{DokoV16}.

 \subsubsection{FSL proof rules.}

\begin{figure}[t]
\begin{center}
\scalebox{0.9}{
\[
\begin{array}{c}
\Inf{\triple{A}{\FenceRel}{\UP{A}}}\quad\quad\Inf{\triple{\DOWN{A}}{\FenceAcq}{A}}\\[3mm]
\Inf{\triple{\UP{\Q(e)} \ast \Rel{l,\Q}}{\RelaxedWrite{l}{e}}{\Init{l}\ast \Rel{l,\Q}}}\\[3mm]
\Inf{\triple{\Init{l} \ast \Acq{l,\Q}}{\RelaxedRead{x}{l}}{\DOWN{\instantiate{\Q}{x}}\ast \Acq{l,\vvar \neq x \Rightarrow \Q}}}\\[5mm]
(A_1 \Rightarrow A_2) \Leftrightarrow (\UP{A_1} \Rightarrow \UP{A_2}) \Leftrightarrow (\DOWN{A_1} \Rightarrow \DOWN{A_2})\\[3mm]
\DOWN(A_1\ast A_2) \equiv (\DOWN{A_1}) \ast (\DOWN{A_2}) \textit{ and analogously for }$\UP$\textit{ and other binary connectives}
\end{array}
\]
}
\end{center}
\caption{Adapted FSL rules for relaxed atomics and fences.}
\label{fig:relaxed-fences-RSL}
\end{figure}

FSL and \FSLpp{} provide proof rules for fences (see \figref{relaxed-fences-RSL}). They use \emph{modalities} $\UP$ (``up'')
and $\DOWN$ (``down'') to represent resources that are transferred
through relaxed accesses and fences.  An assertion $\UP{A}$ represents a
resource $A$ which has been prepared, via a release fence, to be
transferred by a relaxed write operation; dually, $\DOWN{A}$
represents resources $A$ obtained via a relaxed read, which may not be
made use of until an acquire fence is encountered. The proof rule for
relaxed write is identical to that for a release write
(cf.~\figref{release-acquire-RSL}), except that the assertion to be
transferred away must be under the $\UP$ modality; this can be
achieved by the rule for release fences.  The rule for a relaxed read
is the same as that for acquire reads, except that the gained
assertion is under the $\DOWN$ modality. The modality can be removed
by a subsequent acquire fence. Finally, assertions may be rewritten
under modalities, and both modalities distribute over all other
logical connectives.

\figref{FSL-Figure2} shows an example program, which is a variant of the message-passing example from \figref{release-acquire-example-RSL}. Comparing the left-hand one of the three parallel threads, a relaxed read is used in the spin loop; after the loop, this thread will hold the assertion $\DOWN{\pointsto{a}{42}{1}}$. The subsequent $\FenceAcq$ statement allows the modality to be removed, allowing the non-atomic location $a$ to be accessed. Dually, the middle thread employs a $\FenceRel$ statement to place the ownership of the non-atomic locations under the $\UP$ modality, in preparation for the relaxed write to $l$.

\begin{figure}[t]
\begin{center}
\scalebox{0.9}{
\[
\begin{array}{c}
\blue{\Q_1\ \equiv\ (\vvar \neq 0 \Rightarrow \pointsto{a}{42}{1})\quad\quad \Q_2\ \equiv\ (\vvar \neq 0 \Rightarrow \pointsto{b}{7}{1})}\\[5pt]
\blue{\{\truesym\}}\\ 
\NonAtomicAlloc{a};\ \NonAtomicAlloc{b};
\AtomicAllocAcq{l}{\blue{\Q_1{\ast}\Q_2}};\ReleaseWrite{l}{0}\\
\begin{array}{l}
\blue{\{\Acq{l,\Q_1}\ast\Init{l}\}}\\
\slightindent\while{[l]_{\rlx} == 0};\\
\slightindent\FenceAcq;\\
\slightindent\NonAtomicRead{x}{a}\\
\slightindent\NonAtomicWrite{a}{x + 1}\\
\blue{\{\truesym\ast \pointsto{a}{43}{1}\}}
\end{array}
\left|
\left|\
\begin{array}{l}
\blue{\{\Uninit{a}\ast\Uninit{b}\ast\Rel{x,\Q_1{\ast}\Q_2}\}}\\
\slightindent\NonAtomicWrite{a}{42};\\
\slightindent\NonAtomicWrite{b}{7};\\
\slightindent\FenceRel{\blue{\pointsto{a}{42}{1}\ast\pointsto{b}{7}{1}}};\\
\slightindent\RelaxedWrite{l}{1};\\
\blue{\{\truesym\ast\Init{l}\}}
\end{array}
\right|
\right|\
\begin{array}{l}
\blue{\{\Acq{l,\Q_2}\ast\Init{l}\}}\\
\slightindent\while{[l]_{\rlx} == 0};\\
\slightindent\FenceAcq;\\
\slightindent\NonAtomicRead{y}{b};\\
\slightindent\NonAtomicWrite{b}{y + 1}\\
\blue{\{\truesym\ast \pointsto{b}{8}{1}\}}
\end{array}\\
\blue{\{\truesym \ast \pointsto{a}{43}{1} \ast \pointsto{b}{8}{1}\}}
\end{array}
\]}
\end{center}
\myvspace{-4mm}
\caption{A variant of the message-passing example of \figref{release-acquire-example-RSL}, combining relaxed memory accesses and fences to achieve ownership transfer. The example is also a variant of Fig.~2 of the FSL paper \protect\cite{DokoV16}, which is included in our evaluation (\code{FencesDblMsgPass}) in \secref{examples}.}
\label{fig:FSL-Figure2}
\end{figure}

\subsubsection{Encoding.}
\begin{figure}[t]
\begin{silver}[mathescape=true]
domain threeHeaps {
  function up(x: Ref) : Ref;    function upInv(x: Ref) : Ref;
  function down(x: Ref) : Ref;  function downInv(x: Ref) : Ref;
  function heap(x: Ref) : Int;  // identifies which heap a Ref is from
  axiom { forall r:Ref :: upInv(up(r)) == r &&
    (heap(r) == 0 ==> heap(up(r)) == 1 }
  axiom { forall r:Ref :: up(upInv(r)) == r &&
    (heap(r) == 1 ==> heap(upInv(r)) == 0 }
  axiom { forall r:Ref :: downInv(down(r)) == r &&
    (heap(r) == 0 ==> heap(down(r)) == -1 }
  axiom { forall r:Ref :: down(downInv(r)) == r &&
    (heap(r) == -1 ==> heap(downInv(r)) == 0 }
}
$\encodeAss{\UP{A}} \leadsto \upAss{\encodeAss{A}}$ ${}$ ${}$ ${}$ ${}$  $\encodeAss{\DOWN{A}} \leadsto \downAss{\encodeAss{A}}$

$\encodeStm{\RelaxedWrite{l}{e}} \leadsto 
\ldots\textit{encoded as for release writes (\figref{release-acquire-Viper}) except}$
                              $\textit{ using }\upAss{\invariant{i}}\textit{ in place of }\invariant{i}$
$\encodeStm{\RelaxedRead{x}{l}}\leadsto 
\ldots\textit{encoded as for acquire reads (\figref{release-acquire-Viper}) except}$
                              $\textit{ using }\downAss{\invariant{i}}\textit{ in place of }\invariant{i}$

$\encodeStm{\FenceRel{A}}\leadsto 
$ exhale $\encodeAss{A}$; inhale $\upAss{\encodeAss{A}}$

$\encodeStm{\FenceAcq}\leadsto 
$ var rs : Set[Ref]; ${}$ rs := havoc() // unknown set of Refs
  assume forall r: Ref :: r in rs <==> perm(down(r).val) > none
  inhale forall r: Ref :: r in rs ==> acc(r.val, perm(down(r).val))
  assume forall r: Ref :: r in rs ==> r.val == down(r).val
  exhale forall r: Ref :: r in rs ==> acc(down(r).val, perm(down(r).val))
  $\textit{// analogously for each other field, predicate (in place of }$val$\textit{)}$
\end{silver}
\myvspace{-0.8em}
\caption{Viper encoding of the FSL rules for relaxed atomics and memory fences
from \figref{relaxed-fences-RSL}.
We omit triggers for the quantifiers for simplicity, but see~\cite{OnlineAppendix}.}
\label{fig:relaxed-fences-Viper}
\end{figure}

The main challenge in encoding the FSL rules for fences is how to
represent the two new modalities. Since these modalities guard
assertions which cannot be currently used or combined with
modality-free assertions, we model them using two
\emph{additional heaps} to represent the assertions under each modality. The
program heap (along with associated permissions) is a built-in notion
in Viper, and so we cannot directly employ three heaps. Therefore,
we construct the additional ``up'' and
``down'' heaps, by axiomatising bijective mappings $\up$ and $\down$
between a real program reference and its counterparts in these
heaps. That is, technically our encoding represents each source location through
three references in Viper's heap (rather than one reference in three heaps).
Assertions $\UP{A}$ are then represented by replacing \emph{all
  references} $r$ in the encoded assertion $A$ with their counterpart
$\up{r}$. We write $\upAss{A}$ for the transformation which performs
this replacement. For example,
$\upAss{\safecode{acc(x.val) && x.val == 4}}
\ \leadsto\ \safecode{acc(up(x).val) && up(x).val == 4}$. We write
$\downAss{A}$ for the analogous transformation for the $\down$
function.

The extension of our encoding is shown in
\figref{relaxed-fences-Viper}. We employ a Viper \emph{domain} to
introduce and axiomatise the mathematical functions for
our $\up$ and $\down$ mappings. By axiomatising inverses for these
mappings, we guarantee bijectivity. Bijectivity allows Viper to conclude that
(dis)equalities and other information is preserved under these mappings.
Consequently, we do not have to explicitly encode the last two rules
of \figref{relaxed-fences-RSL}; they are reduced to standard assertion
manipulations in our encoding.
An additional $\heap$ function labels references with an integer
identifying the heap to which they belong (\code{0} for real
references, \code{-1} and \code{1} for their ``down'' and ``up''
counterparts); this labelling provides the verifiers
with the (important) information that these notional heaps are
disjoint.

Our handling of relaxed reads and writes
is almost identical to that of acquire reads and release
writes in \figref{release-acquire-Viper}; this similarity comes from
the proof rules, which only require that the modalities be
inserted for the invariant. Our
encoding for release fences requires an annotation in the source
program to indicate which assertion to prepare for release by
placing it under the $\UP$ modality.

\begin{figure}[t]
\begin{center}
\scalebox{0.9}{
\[
\begin{array}{c}
\Inf{\triple{\truesym}{\AtomicAllocRMW{l}{\Q}}{\Rel{l,\Q} \ast \RMWAcq{l,\Q}}}\\[4mm]
\Inf{\begin{array}{l}
x\notin\FV{P}\\
x\notin\FV{e}\end{array}\quad}{\quad P'\equiv\left\{\begin{array}{ll}
P\quad&\textit{if }\tau\in\{\rel,\relacq\}\\
\UP{P}\quad&\textit{otherwise}\end{array}
\right.}.{\begin{array}{l}
\instantiate{\Q}{e}\entails A\ast T\\
P\ast T \entails \instantiate{\Q}{e'}\end{array}}{ A'\equiv\left\{\begin{array}{ll}
A\quad&\textit{if }\tau\in\{\acq,\relacq\}\\
\DOWN{A}\quad&\textit{otherwise}\end{array}
\right.}{\doubleRowTriple{\Init{l} \ast \Rel{l,\Q} \ast\ }{\RMWAcq{l,\Q} \ast P'}{\CAS{x}{\tau}{l}{e}{e'}}{\cond{x=e}{A'}{P'} \ast \Init{l}\ \ast}{\Rel{l,\Q} \ast \RMWAcq{l,\Q}}}\\
\ \\[-2mm]
\RMWAcq{l,\Q}  \quad\Leftrightarrow\quad  \RMWAcq{l,\Q} \ast \RMWAcq{l,\Q}
\end{array}
\]}
\end{center}
\myvspace{-4mm}
\caption{Adapted \FSLpp{} rules for compare and swap operations.
\textit{FV} yields the free variables of an assertion.}
\label{fig:compare-and-swap-RSL}
\end{figure}

Our encoding for acquire fences
does \emph{not} require any annotations.  \emph{Any} assertion
under the $\DOWN$ modality can (and should) be converted to
its corresponding version without the modality because $\DOWN{A}$ is
strictly less-useful than $A$ itself. To encode this conversion, we
find \emph{all} permissions currently held in the down
heap, and transfer these permissions and the values of the
corresponding locations over to the real heap. These steps are encoded
for each field and
predicate separately; \figref{relaxed-fences-Viper} shows the steps for the \code{val} field. We first define a set
\code{rs} to be precisely the set of all references \code{r} to which
\emph{some} permission to \code{down(r).val} is
currently held, \ie{}, \code{perm(down(r).val) > none}. For each
such reference, we \code{inhale} exactly the same amount of permission
to the corresponding \code{r.val} location, equate the heap values,
and then remove the permission to the down locations.

With our encoding based on multiple heaps, reasoning about assertions
under modalities  inherits all of Viper's native
automation for permission and heap reasoning.  We
will reuse this idea for a different purpose in the following section.

\section{Compare and Swap}
\label{sec:compare-and-swap}

C11 includes atomic \emph{read-modify-write} operations, commonly used
to implement high-level synchronisation primitives such as locks.
\FSLpp{} \cite{DokoVafeiadis17} provides
proof rules for \emph{compare and swap} (CAS) operations. An atomic
compare and swap $\CAS{\tau}{l}{e}{e'}$ reads and returns the value of
location $l$; if the value read is equal to $e$, it also writes the
value $e'$ (otherwise we say that the CAS \emph{fails}).

\subsubsection{$\FSLpp$ proof rules.}
$\FSLpp$
provides an assertion
$\RMWAcq{l,\Q}$, which is similar to $\Acq{l, \Q}$, but is used for
CAS operations instead of acquire
reads. A successful CAS
\emph{both} obtains ownership of an assertion via its read operation
and gives up ownership of an assertion via its write operation.

\FSLpp{} does not support general combinations of atomic reads and
CAS operations on the same location; the way of reading must be chosen
at allocation via the annotation $\rho$ on the allocation statement
(see \figref{syntaxone}). In contrast to the $\AcqX$ assertions used
for atomic reads, $\RMWAcqX$ assertions can be freely duplicated and
their invariants need not be adjusted for a successful CAS: when using only CAS operations, each value read from a location corresponds to a
different write.

Our presentation of the relevant proof rules is shown in
\figref{compare-and-swap-RSL}. Allocating a location with annotation
$\rmw$ provides a $\RelX$ and a $\RMWAcqX$ assertion, such that
the location can be used for release writes and CAS operations.

For the CAS operation, we present a single, general proof rule instead
of four rules for the
different combinations of access modes in \FSLpp{}.
The rule requires that $l$ is initialised (since its
value is read), $\RelX$ and $\RMWAcqX$ assertions,
and an assertion $P'$ that provides the resources needed for a successful
CAS\@. If the CAS fails (that is, $x\not= e$), its precondition is
preserved.

\begin{figure}[t]
\begin{center}
(i)\includegraphics[height=2.7cm]{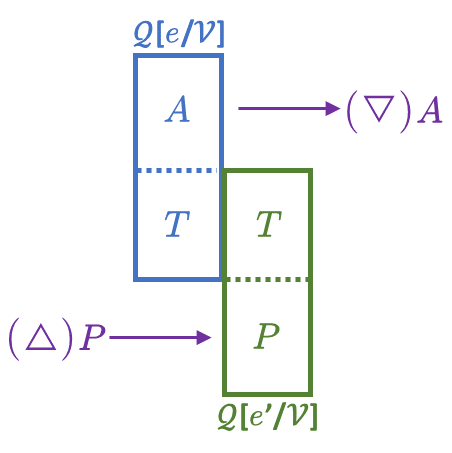}\quad\quad(ii)\includegraphics[height=2.7cm]{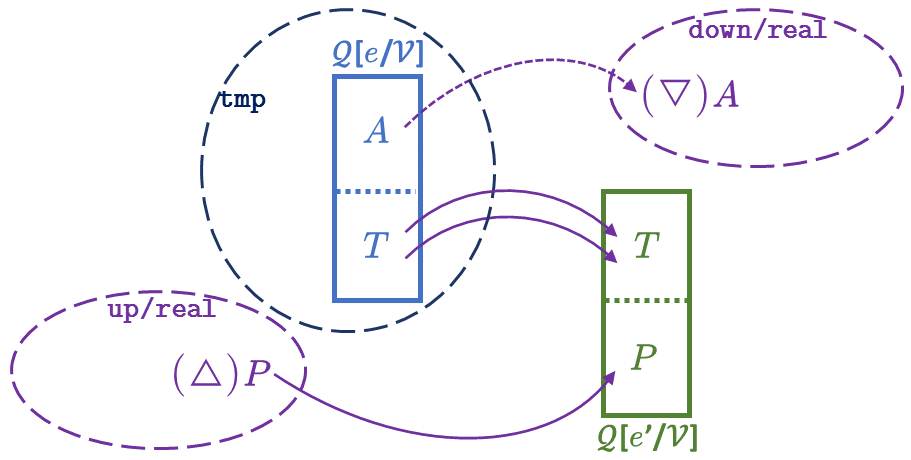}
\end{center}
\myvspace{-4mm}
\caption{An illustration of (i) the proof rule for CAS operations and
(ii) our Viper encoding; the dashed regions denote the relevant heaps employed in the encoding.}
\label{fig:compare-and-swap-diagrams}
\end{figure}

If the CAS succeeds, it has read value $e$ and written value $e'$.
Assuming for now that the access mode $\tau$ permits ownership transfer,
the thread has acquired $\instantiate{\Q}{e}$ and released
$\instantiate{\Q}{e'}$. As illustrated in \figref{compare-and-swap-diagrams}(i),
these assertions may overlap. Let $T$ denote the assertion characterizing the
overlap; then assertion $A$ denotes $\instantiate{\Q}{e}$ without the overlap, and
$P$ denotes $\instantiate{\Q}{e'}$ without the overlap.
The net effect of a successful CAS is then to acquire $A$ and to release
$P$, while $T$ remains with the location invariant across the CAS\@.
Automating the choice of $T$, $A$, and $P$ is one
of the main challenges of encoding this rule.
Finally, if the access mode $\tau$ does not permit ownership transfer
(that is, fences are needed to perform the transfer), $A$ and $P$
are put under the appropriate modalities.

\subsubsection{Encoding.}
Our encoding of CAS operations uses several techniques presented in
earlier sections: see \appref{compare-and-swap-details} for details.  We
represent $\RMWAcqX$ assertions analogously to our encoding of $\AcqX$
assertions (see \secref{release-acquire}). We use the value of
field \code{acq} (cf.~\figref{release-acquire-Viper}) to distinguish holding some $\RMWAcqX$ assertion from some $\AcqX$
assertion. Since $\RMWAcqX$ assertions are duplicable
(cf.~\figref{compare-and-swap-RSL}), we employ \code{wildcard}
permissions for the corresponding \code{AcqConjunct} predicates.

Our encoding of the proof rule for CAS operations is somewhat
involved; we give a high-level description here, and
relegate the details to \apprefnocite{compare-and-swap-details}. We focus on the
more-interesting case of a successful CAS here. The key
challenge is how to select assertion $T$ to satisfy the
premises of the rule. Maximising this overlap is
desirable in practice since this reduces the resources to be
transferred, and which must interact in some cases with the modalities.
Our Viper encoding indirectly \emph{computes}
this largest-possible $T$ as follows (see
\figref{compare-and-swap-diagrams}(ii) for an illustration).

We introduce yet another heap
(``\tempName'') in which we inhale the invariant $\instantiate{\Q}{e}$ for the value
read. Now, we exhale the
invariant $\instantiate{\Q}{e'}$ for the value written, but adapt
the assertions as follows: for each permission in the invariant, we
take \emph{the maximum possible} amount from our ``\tempName'' heap;
these permissions correspond to $T$.
Any remainder is taken from the current heap (either the real or the
``up'' heap, depending on $\tau$); these correspond to $P$.
Any permissions remaining in the
``\tempName'' heap after this exhale correspond to the assertion $A$ and are moved (in a way
similar to our \FenceAcq{}
encoding in \figref{relaxed-fences-Viper}) to either the real or
``\downName'' heap (depending on $\tau$).

This combination of techniques results in an automatic support for the proof
rule for CAS statements.
This completes the core of our Viper encoding, which now handles the complete set of memory access constructs from \figref{syntaxone}.

\section{Soundness and Completeness}
\label{sec:soundness}

We give a brief overview of the soundness argument for our encoding here, and also discuss where it can be incomplete compared with a manual proof effort; further details are included in \appref{soundnessdetails}.

\subsubsection{Soundness.}
Soundness means that if the Viper encoding of a program and its
  specification verifies, then there exists a proof of the program
and specification using the RSL logics. We can show this property in two
main steps.
First, we show that the states before and after each encoded statement
in the Viper program satisfy several invariants.  For example, we
never hold permissions to a non-atomic reference's \code{val} field
but not its \code{init} field.
Second, we reproduce a Hoare-style proof outline in the RSL logics.  For
this purpose, we define a mapping from \emph{states} of the Viper
program back to RSL \emph{assertions} and show two properties:
(1)~When we map the initial and final states of an encoded program
statement to RSL assertions, we obtain a provable Hoare triple.
(2)~Any automatic entailment reasoning performed by Viper coincides
with entailments sound in the RSL logics.
These two facts together
imply that our technique will only verify (encoded)
properties for which a proof exists in the RSL logics; \ie{} our
technique is sound.

\subsubsection{Completeness.}

Completeness means that all programs provable in the RSL logics can
be verified via their encoding into Viper. By systematically analysing each
rule of these logics, we identify three sources of incompleteness of our encoding:
(1)~It does not allow one to strengthen the invariant in a $\RelX{}$
assertion; strengthening the requirement on writing does not allow more programs to be verified \cite{ViktorPersonalCommunication}.
(2)~For a $\FenceAcq$, our encoding removes \emph{all} assertions from under a $\DOWN$ modality. As explained in \secref{fences}, the ability to choose \emph{not} to remove the modality is not useful in practice.
(3)~The ghost state employed in \FSLpp{} can be defined over a \emph{custom permission structure} (partial commutative monoid), which is not possible in Viper.
This is the only incompleteness of our encoding arising in practice; we will
discuss an example in \secref{examples}.


\section{Examples and Evaluation}
\label{sec:examples}
\begin{figure}[t]
\setlength{\tabcolsep}{4pt}
\newcolumntype{d}[1]{D{.}{.}{#1}}
\newcommand{\h}[1]{\multicolumn{1}{c}{#1}}
\newcommand{\hb}[1]{\multicolumn{1}{c|}{#1}}
\begin{center}
\scalebox{0.8}{
\begin{tabular}{l|c|c|r|c|c|c|r} 
  Program  & \hb{Prototype} & \hb{Size (LOC,}  &\hb{Time} & \multicolumn{2}{c|}{Specs} & \hb{Other} & \h{Coq} \\
          & \hb{support} & \hb{funcs,loops)} &\hb{(s)} & \hb{PP} & \hb{LI}          & \hb{Annot.} & \h{Annot.}\\
  \hline
  \texttt{RSLSpinLock}             & \tick  & 7,3,2 & 10.83 & 3 & 1 & 1 & 120 \cite{VafeiadisN13} \\
  \texttt{RSLLockNoSpin}           & \tick  & 6,3,1 & 10.33 & 3 & 0 & 1 & 84 \cite{Kaiser17}  \\
  \texttt{RSLLockNoSpin\_err}      & \tick   & 6,3,1 & 9.74 & 3 & 0 & 1 & \notapplicable  \\
  \texttt{RelAcqMsgPass}           & \tick  & 15,3,1 & 10.46 & 3 & 0 & 1 & 99 \cite{VafeiadisN13} \\
  \texttt{RelAcqMsgPass\_err}   & \tick & 15,3,1 & 9.57 & 3 & 0 & 1 & \notapplicable \\
  \texttt{RelAcqDblMsgPassSplit}   & \tick       & 21,4,2 & 10.84 & 4 & 0 & 1 & \notapplicable \\
  \texttt{RelAcqDblMsgPassSplit\_err}& \tick& 21,4,2 & 9.86 & 4 & 0 & 1 & \notapplicable \\
  \texttt{CASModesTest}             & \tick& 23,3,2 & 18.05 & 3 & 0 & 2 & \notapplicable  \\
  \texttt{CASModesTest\_err}        & \tick     & 24,3,2 & 17.50 & 3 & 0 & 2 & \notapplicable  \\
  \texttt{FencesDblMsgPass}         & \tick    & 27,4,2 & 12.32 & 4 & 0 & 3 & \notapplicable  \\
  \texttt{FencesDblMsgPass\_err}    & \tick         & 27,4,2 & 10.73 & 4 & 0 & 3 & \notapplicable  \\
  \texttt{FencesDblMsgPassSplit}      & \tick       & 24,4,2 & 12.61 & 4 & 0 & 2 & \notapplicable  \\
  \texttt{FencesDblMsgPassSplit\_err}     & \tick        & 24,4,2 & 11.53 & 4 & 0 & 2 & \notapplicable  \\
  \texttt{FencesDblMsgPassAcqRewrite}    &         & 24,4,2 & 15.75 & 4 & 0 & 3 & \notapplicable  \\
  \texttt{RustARCOriginal\_err}           &  & 10,4,0 & 37.53 & 4 & 0 & 2 & 654 \cite{DokoVafeiadis17}  \\
  \texttt{RustARCStronger}           &  & 10,4,0 & 31.86 & 4 & 0 & 2 & \notapplicable  \\
  \texttt{RelAcqRustARCStronger}      &       & 9,4,0 & 15.75 & 4 & 0 & 2 & \notapplicable  \\
  \texttt{FollyRWSpinlock\_err}        &     & 24,7,2 & 28.21 & 7 & 2 & 0 & \notapplicable  \\
  \texttt{FollyRWSpinlockStronger}      &       & 26,7,3 & 21.93 & 7 & 3 & 0 & \notapplicable  \\
\end{tabular}
}
\end{center}
\myvspace{-1.0em}
\caption{The results of our evaluation. Examples including \texttt{\_err} are expected to generate errors; those with \texttt{Stronger} are variants of the original code with less-efficient atomics and a correspondingly different proof. Under ``Size'', we measure lines of code, number of distinct functions/threads, and number of loops. Under ``Specs'', ``PP'' stands for the necessary pairs of pre and post-conditions; ``LI'' stands for loop invariants required. ``Other Annot.'' counts any other annotations needed.
For examples that have been verified in Coq,
we report the number of manual proof steps (in addition to pre-post pairs)
and provide a reference to the proof.}
\label{fig:experiments}
\end{figure}

We evaluated our work with a prototype front-end tool
 \cite{PrototypeTool}, 
and some additional
experiments directly at the Viper level~\cite{OnlineAppendix}.
Our front-end tool accepts a simple
input language for C11 programs, closely modelled on the syntax of the
RSL logics. It supports all  features described in this
paper, with the exception of invariant rewriting (\cf{}
\appref{rewriting-invariants}) and ghost state (\apprefnocite{ghosts}), which will be simple extensions. We
encoded examples which require these features, additional
theories, or custom permission structures manually into Viper, to simulate
what an extended version of our prototype
will be able to achieve.

Our encoding supports several extra features which we used in our
experiments but mention only briefly here: (1)~We support the \FSLpp{}
rules for \emph{ghost state}: see \apprefnocite{ghosts}. (2)~Our encoding
handles common spin loop patterns without requiring loop invariant annotations. (3)~We support fetch-update instructions (\eg{} atomic
increments) natively, modelled as a CAS which never fails.

\noindent
\textbf{Examples.}
We took examples from the RSL \cite{VafeiadisN13} and FSL \cite{DokoV16}
papers, along with variants in which we seeded errors, to
check that verification fails as expected (and in comparable time).
We also encoded the Rust
reference-counting (ARC) library~\cite{RustReferenceCounting}, which is
the main example from \FSLpp{}~\cite{DokoVafeiadis17}. The
proof there employs a custom permission structure,
which is not yet supported by Viper.
However, following the suggestion of one of the authors
\cite{ViktorPersonalCommunication}, we were able to fully verify two
variants of the example, in which some access modes are
strengthened, making the code slightly less efficient but enabling a
proof using a simpler permission model. For these variants, we
required \emph{counting permissions}~\cite{Bornat05}, which we
expressed with additional background definitions
(see \cite{OnlineAppendix} for details, and \appref{example-details} for the code).
Finally, we tackled seven core functions of a reader-writer-spinlock from the Facebook Folly
library \cite{FollyRWSpinLock}.
We were able to
verify five of them directly. The other two employ code idioms which seem to be beyond
the scope of the RSL logics, at least without sophisticated
ghost state. For both functions, we also wrote and verified alternative
implementations.
The
Rust and Facebook examples demonstrate a key advantage of
building on top of Viper; both
require support for extra theories (counting permissions as well as
modulo and bitwise arithmetic), which we were able to encode easily.

\noindent
\textbf{Performance.}
We measured the verification times on an Intel Core i7-4770 CPU
(3.40GHz, 16Gb RAM) running Windows 10 Pro and report the average
of 5 runs. For those examples
supported by our front-end, the times include the generation of the
Viper code.  As shown in \figref{experiments}, verification times are
reasonable (generally around 10 seconds, and always under a minute).

\noindent
\textbf{Automation.}
Each function (and thread) must be annotated with an appropriate pre
and post-condition, as is standard for modular verification.  In
addition, some of our examples require loop invariants and
other annotations (\eg{} on allocation statements).  Critically, the
number of such annotations is very low.  In particular, our annotation
overhead is between one and two orders of magnitude lower than the
overhead of existing mechanised proofs (using the Coq formalisations
for \cite{VafeiadisN13,DokoVafeiadis17} and a recent encoding
\cite{Kaiser17} of RSL into Iris \cite{Krebbers17}).
Such ratios are consistent
with other recent Coq-mechanised proofs based on separation logic~(\eg{} \cite{Xu2016}), which suggests that the strong soundness guarantees provided by Coq
have a high cost when \emph{applying} the logics.
By contrast, once the specifications are provided, our approach is almost entirely automatic.

\section{Conclusions and Future Work}
\label{sec:conclusion}

We have presented the first encoding of modern program logics for weak
memory models into an automated deductive program
verifier. The encoding enables programs (with suitable annotations) to
be verified automatically by existing back-end tools.
We have implemented a
front-end verifier and demonstrated that our encoding can be used to
verify weak-memory programs efficiently and with low annotation overhead.
As future work, we plan to tackle other weak-memory logics such as GPS
\cite{TuronVD14}.  Building practical tools that implement such
advanced formalisms will provide feedback that inspires further
improvements of the logics.


%
%
%
%
%
%
%
%
%

\subsubsection{Data Availability Statement and Acknowledgements.}
The artifact accompanying our submission is available in the TACAS figshare repository \cite{PrototypeTool} at \url{\artifacturl{}}

We are grateful to Viktor
Vafeiadis and Marko Doko for many explanations of the RSL logics and helpful
discussions about our encoding. We thank Christiane Goltz for her work on the prototype tool, and Malte Schwerhoff for implementing additional features. We thank Marco Eilers for his assistance with the online appendix, and Arshavir Ter-Gabrielyan for automating our artifact assembly for various operating systems.
We also thank Andrei Dan, Lucas Brutschy and Malte
Schwerhoff for feedback on earlier versions of this manuscript.

\ifx\istr\undefined
\newpage
\fi

\bibliographystyle{abbrv}
\bibliography{RSLtoViper}

\ifx\istr\undefined
\newpage
\fi

\begin{appendix}

\comment{
Left-over:
 In addition, just as for
classical Hoare Logic proofs, many steps of these arguments can
typically be automated, leaving the user to provide only the
``essential'' details of the argument (analogous to loop invariants in
sequential programs).
}

\section{Full Source Encoding}
\label{sec:source-encoding}

The encoding of the general source language of assertions is given below (we assume the encoding of pure expressions, which can typically be the identity mapping, assuming all operators supported such as addition, equality etc.{} are all supported natively by Viper).
\[
\begin{array}{rcll}
\encodeAss{\emp} &\leadsto& \truesym\\
\encodeAss{\pointsto{l}{e}{k}} &\leadsto& \texttt{acc(l.val, k) \&\& acc(l.init, k) \&\&}\\
\multicolumn{3}{r}{\texttt{l.val==}\encodeAss{e}\texttt{ \&\& l.init}}\\
\encodeAss{A_1 \ast A_2} &\leadsto& \encodeAss{A_1} \texttt{\&\&} \encodeAss{A_2}\\
\encodeAss{b\Rightarrow A} &\leadsto& \encodeAss{b}\Rightarrow \encodeAss{A}\\
\encodeAss{\cond{b}{A_1}{A_2}} &\leadsto& \cond{\encodeAss{b}}{\encodeAss{A_1}}{\encodeAss{A_2}}\\
\encodeAss{\Uninit{l}}&\leadsto& \texttt{acc(l.val) \&\& acc(l.init) \&\& !l.init}\\
\encodeAss{\Acq{l,\Q}}&\leadsto& \texttt{acc(l.acq, wildcard) \&\& l.acq==true \&\& }\\
\multicolumn{3}{r}{(\foreach{i}{\conjuncts{\Q}}\texttt{: AcqConjunct(l, }i\texttt{) \&\&\ \ }}\\
\multicolumn{3}{r}{\texttt{valsRead(l, }i\texttt{)==Set[Int]() }\foreachEnd)}\\
\encodeAss{\Rel{l,\Q}}&\leadsto& \texttt{acc(l.rel, wildcard) \&\& l.rel==}\whole{\Q}\\
\encodeAss{\Init{l}}&\leadsto& \texttt{acc(l.init, wildcard)}\\
\encodeAss{\UP{A}}&\leadsto& \upAss{\encodeAss{A}}\\
\encodeAss{\DOWN{A}}&\leadsto& \downAss{\encodeAss{A}}\\
\encodeAss{\RMWAcq{l,\Q}}&\leadsto& \texttt{acc(l.acq, wildcard) \&\& l.acq==false \&\& }\\
\multicolumn{3}{r}{(\foreach{i}{\conjuncts{\Q}}\texttt{: acc(AcqConjunct(l, }i\texttt{), wildcard) } \foreachEnd)}\\
\end{array}
\]
For example which potentially employ multiple copies of the same conjunct in an $\Acq$ predicate's invariant, some additional care needs to be taken about when exactly to make the $\texttt{valsRead(l, }i\texttt{)}==Set[Int]()$ assumption; this is discussed in \appref{multipleconjuncts}.

\section{Example Details}
\label{sec:example-details}

To give an impression of the input required for our encoding, we provide source code corresponding to some of our encoded examples from the Online Appendix \cite{OnlineAppendix}, given in Figures \ref{fig:RSL-Figure7} to \ref{fig:last}. For supported examples (\cf{} \figref{experiments}), one can also see the input files for our prototype tool \cite{PrototypeTool}.

\begin{figure}
\[
\begin{array}{l}
\blue{\Q\ \equiv\ \cond{\vvar = 0}{\truesym}{\cond{\vvar = 1}{J}{\falsesym}}}\\[5pt]
\blue{\textit{Lock}(x)\ \equiv\ \Init{x}\ast\RMWAcq{x,\Q}\ast\Rel{x,\Q}}\\[5pt]
\begin{array}{ll}
\begin{array}{l}
\texttt{new\_lock() returns (x)}\\
\blue{\textit{\ \ requires }J}\\
\blue{\textit{\ \ ensures }\textit{Lock}(x)}\\
\texttt{\{}\\
\quad\AtomicAllocAcq{x}{\blue{\Q}};\\
\quad\ReleaseWrite{x}{1};\\
\texttt{\}}\\
\end{array}
\quad\quad&\quad\quad
\begin{array}{l}
\texttt{unlock(x)}\\
\blue{\textit{\ \ requires }J\ast\textit{Lock}(x)}\\
\blue{\textit{\ \ ensures }\textit{Lock}(x)}\\
\texttt{\{}\\
\quad\ReleaseWrite{x}{1};\\
\texttt{\}}\\
\end{array}\\
\ \\
\texttt{lock(x)}\\
\blue{\textit{\ \ requires }\textit{Lock}(x)}\\
\blue{\textit{\ \ ensures }J\ast\textit{Lock}(x)}\\
\texttt{\{}\\
\quad\while{\CAS{\relacq}{x}{1}{0}\texttt{ != 1}};\\
\texttt{\}}\\
\end{array}
\end{array}
\]
\vspace{-0.75em}
\caption{\texttt{RSLLockNoSpin} example (based on RSL Figure 7). Annotations in \blue{blue}.}
\label{fig:RSL-Figure7}
\end{figure}

%

\begin{figure}[t]
\[
\begin{array}{c}
\blue{\Q_1\ \equiv\ (\vvar \neq 0 \Rightarrow \pointsto{a}{42}{1})\quad\quad \Q_2\ \equiv\ (\vvar \neq 0 \Rightarrow \pointsto{b}{7}{1})}\\
\blue{\Q_3\ \equiv\ (\vvar \neq 0 \Rightarrow \pointsto{a}{42}{1} \ast \pointsto{b}{7}{1})}\\[5pt]
\blue{\{\truesym\}}\\ 
\NonAtomicAlloc{a};\ \NonAtomicAlloc{b};
\AtomicAllocAcq{x}{\blue{\Q_3}};\\
\blue{\Rewrite{\Acq{x,\Q_3}}{\Acq{x,\Q_1\ast\Q_2}}};\\
\ReleaseWrite{x}{0};\\
\begin{array}{l}
\blue{\{\Acq{x,\Q_1}\ast\Init{x}\}}\\
\slightindent\while{[x]_{\rlx} == 0};\\
\slightindent\FenceAcq;\\
\slightindent\NonAtomicRead{z}{a}\\
\slightindent\NonAtomicWrite{a}{z + 1}\\
\blue{\{\truesym\ast \pointsto{a}{43}{1}\}}
\end{array}
\left|
\left|\
\begin{array}{l}
\blue{\{\Uninit{a}\ast\Uninit{b}\ast\Rel{x,\Q_3}\}}\\
\slightindent\NonAtomicWrite{a}{42};\\
\slightindent\NonAtomicWrite{b}{7};\\
\slightindent\FenceRel{\blue{\pointsto{a}{42}{1}\ast\pointsto{b}{7}{1}}};\\
\slightindent\RelaxedWrite{x}{1};\\
\blue{\{\truesym\ast\Init{x}\}}
\end{array}
\right|
\right|\
\begin{array}{l}
\blue{\{\Acq{x,\Q_2}\ast\Init{x}\}}\\
\slightindent\while{[x]_{\rlx} == 0};\\
\slightindent\FenceAcq;\\
\slightindent\NonAtomicRead{w}{b};\\
\slightindent\NonAtomicWrite{b}{w + 1}\\
\blue{\{\truesym\ast \pointsto{b}{8}{1}\}}
\end{array}\\
\blue{\{\truesym \ast \pointsto{a}{43}{1} \ast \pointsto{b}{8}{1}\}}
\end{array}
\]
\vspace{-4mm}
\caption{\texttt{FencesDblMsgPassAcqRewrite} examples. Annotations in \blue{blue}.}
\label{fig:FSLFigure2variantWithRewriting}
\end{figure}

\newcommand{\rd}{\textit{rd}}

\begin{figure}
\[
\begin{array}{l}
\blue{\Q\ \equiv\ \vvar \geq 0 \ast \pointsto{g}{\_}{1 - \vvar*\rd} \ast (\vvar \geq 1 \Rightarrow \pointsto{d}{\_}{1 - \vvar*\rd})}\\[5pt]
\blue{\textit{ARC}(d,c,g,v)\ \equiv\ \pointsto{d}{v}{\rd} \ast \pointsto{g}{\_}{\rd} \ast \RMWAcq{c,\Q} \ast \Rel{c,\Q} \ast \Init{c}}\\[5pt]
\begin{array}{ll}
\begin{array}{l}
\texttt{new(v) returns (d,c,g)}\\
\blue{\textit{\ \ requires }\truesym}\\
\blue{\textit{\ \ ensures }\textit{ARC}(d,c,g,v)}\\
\texttt{\{}\\
\quad\NonAtomicAlloc{d};\\
\quad\blue{\GhostAlloc{g};}\\
\quad\AtomicAllocRMW{c}{\Q};\\
\quad\NonAtomicWrite{d}{v};\\
\quad\ReleaseWrite{c}{1};\\
\texttt{\}}\\
\end{array}
\quad\quad&\quad\quad
\begin{array}{l}
\texttt{drop(d,c,g)}\\
\blue{\textit{\ \ requires }\textit{ARC}(d,c,g,\_)}\\
\blue{\textit{\ \ ensures }\truesym}\\
\texttt{\{}\\
\quad\FetchAdd{t}{\rel}{c}{-1};\\
\quad\texttt{if (}t\texttt{==1)\{}\\
\quad\quad\FenceAcq;\\
\quad\quad\texttt{free(d)};\\
\quad\texttt{\}}\\
\texttt{\}}\\
\end{array}
\ \\
\ \\
\begin{array}{l}
\texttt{read(d,c,g) returns (v)}\\
\blue{\textit{\ \ requires }\textit{ARC}(d,c,g,\_)}\\
\blue{\textit{\ \ ensures }\textit{ARC}(d,c,g,v)}\\
\texttt{\{}\\
\quad\NonAtomicRead{v}{d};\\
\texttt{\}}\\
\end{array}
\quad\quad&\quad\quad
\begin{array}{l}
\texttt{clone(d,c,g)}\\
\blue{\textit{\ \ requires }\textit{ARC}(d,c,g,v)}\\
\blue{\textit{\ \ ensures }\textit{ARC}(d,c,g,v)\ast\textit{ARC}(d,c,g,v)}\\
\texttt{\{}\\
\quad\FetchAddNo{\acq}{c}{1};\\
\texttt{\}}\\
\end{array}
\end{array}
\end{array}
\]
\vspace{-0.75em}
\caption{Rust reference counting variant with strengthened access modes (\texttt{RustARCStronger} in our evaluation). Compared to the original code (see \cite{DokoVafeiadis17}) we modified the write in \texttt{new} to use a release rather than relaxed mode, and the update in \texttt{clone} to use acquire rather than relaxed. As discussed in the paper body, the original version of the example is proved in \cite{DokoVafeiadis17} using features which are not yet supported by our encoding. We do, however, exploit the CAS rules and rules for fences here. We write $\rd$ for a read permission, in the sense of counting permissions \cite{Bornat05}. The ghost location $g$ must be identifiable as such for the encoding, for example by considering this a type annotation, or using a distinguished class of variables for ghost locations. We model the \texttt{free} statement by exhaling the corresponding permissions.
}
\label{fig:last}
\end{figure}
%
%
%
%

\section{Rewriting Invariants}
\label{sec:rewriting-invariants}

\begin{figure}[t]
\begin{silver}[mathescape=true]
$\encodeStm{\Rewrite{\Acq{l,\Q}}{\Acq{l,\Q'}}}\leadsto
$
  assert SomeAcq(l)
  var tmpBool : Bool
  tmpBool := havoc()

  if(tmpBool) { // check rewriting is justified

    // remove all permissions from current state
    exhale forall r: Ref :: r != null ==> acc(r.init, perm(r.init))
    exhale forall r: Ref :: r != null ==> acc(r.val, perm(r.val))
    exhale forall r: Ref :: r != null ==> acc(r.rel, perm(r.rel))
    exhale forall r: Ref :: r != null ==> acc(r.acq, perm(r.acq))
  //$\textit{ analogously for other fields, predicates in source program}$

    var v :Int
    v := havoc() // perform check for arbitrary v

    // inhale original invariant
    $\foreachNumber$ $\foreachStart$
    if(i in $\conjuncts{\Q}$) {
      inhale $\instantiate{\invariant{i}}{v}$
    }
    $\foreachEnd$

    // exhale new invariant
    $\foreachNumber$ $\foreachStart$
    if(i in $\conjuncts{\Q'}$) {
      exhale $\instantiate{\invariant{i}}{v}$
    }
    $\foreachEnd$

    assume false // kill this branch - we've checked rewriting is OK
  }

  // update the conjuncts held
  exhale ($\foreach{i}{\conjuncts{\Q}}$:
  AcqConjunct(l, $i$) && valsRead(l, $i$) == Set[Int]() $\foreachEnd$)
  inhale ($\foreach{i}{\conjuncts{\Q'}}$:
  AcqConjunct(l, $i$) && valsRead(l, $i$) == Set[Int]() $\foreachEnd$)

\end{silver}
\vspace{-0.75em}
\caption{Viper encoding of a source-level \texttt{Rewrite} statement.}
\label{fig:rewriting}
\end{figure}

It is unusual (at least, in the examples we have investigated so far) for very many different invariants for atomic locations to be needed; it is even less common for there to be a need for many different invariants for the \emph{same} atomic location. Indeed, for $\RelX$ and $\RMWAcqX$ assertions, since the assertions are duplicable, one may as well always use the same invariant. For $\AcqX$ assertions the situation is more interesting; it may be desirable to split the invariant (as used e.g.~in \figref{FSLFigure2variantWithRewriting}) across several $\AcqX$ assertions, and programmer-annotated assertions may not always syntactically match up precisely (since there might be more readable ways of expressing an equivalent assertion). Since our indexing of $\AcqX$ invariants matches their conjuncts syntactically, additional work is required if this syntactic match would be overly restrictive. For example, in the example shown in \figref{FSLFigure2variantWithRewriting}, the initial $\AcqX$ invariant is expressed more succinctly in a way which does not provide the immediate conjuncts needed by the (left and right) forked threads. In such cases, we support an additional \code{rewrite} statement $\Rewrite{\Acq{l,\Q}}{\Acq{l,\Q'}}$ in the source program to explain the intended rewriting. To \emph{check} that such a statement is justified, we need to check the entailment between the original and new invariants, for all values of $\vvar$. Furthermore, this entailment check cannot be made directly in the \emph{current} state, since that might already contain permissions and value information which could unsoundly weaken the check made, or even contradict the invariants involved, resulting in an infeasible program state from there onwards.

To avoid these issues, we perform the following steps (shown in \figref{rewriting}). For simplicity, we do not handle the case of rewriting invariants for which values have already been read (we check that this is not the case, here, but an extension is possible). Firstly, we create a non-deterministic \code{if}-branch. Inside the branch we \emph{remove} all permissions from the current state. We then havoc an integer variable, representing the arbitrary value of $\vvar$ for which the subsequent check should hold. We \code{inhale} the original invariant (using our indexing as usual), and \code{exhale} the invariant to which it is to be rewritten. If this succeeds, we know that the rewriting is justified; the one invariant entails the other, for all values of $\vvar$. We then kill this branch, by adding an \code{assume false} to it; subsequently, only the other branch (in which no changes were made) will be considered for verification.

Lastly, we perform the rewriting itself by discarding all of the original \code{AcqConjunct} instances, and replacing them with the new ones. Verification can then proceed as usual.

\subsection{Multiple Copies of Invariant Conjuncts}
\label{sec:multipleconjuncts}
If an example exhibits the following sequence of steps: we inhale an $\Acq{l,\Q}$ conjunct, then we perform an acquire read, and then we inhale (perhaps due to joining a thread) \emph{another} $\Acq{l,\Q}$ conjunct, then the presented encoding of this assertion is not quite correct. The problem is that we must avoid ``re-initialising'' the function tracking the values read in the location; blindly assuming $\texttt{valsRead(l, }i\texttt{)}==Set[Int]()$ (as in \secref{source-encoding}) can lead to contradictions. We solve this problem simply by making the assumption \emph{only} if the newly-acquired conjunct was not already held. This is easy to check using Viper's permission introspection. Note that this comes with a new incompleteness (albeit for an extremely specific situation): we effectively ``obliterate'' one point in the acquire conjunct (for the earlier value read) in \emph{both} copies of the conjunct, where technically we need only do so for the one formerly held. We could extend our modelling to handle this situation, but it seems unnecessary in practice.

\section{Ghost Locations}
\label{sec:ghosts}

\begin{figure}[t]
\begin{silver}[mathescape=true]
define realRef(x) !is_ghost(x) && heap(x) == 0
define ghostRef(x) is_ghost(x) && heap(x) == 0

domain parallelHeaps {
  function up(x: Ref) : Ref
  function down(x: Ref) : Ref
  function up_inv(x: Ref) : Ref
  function down_inv(x: Ref) : Ref

  function temp(x: Ref) : Ref
  function temp_inv(x: Ref) : Ref

  function heap(x: Ref) : Int
  function is_ghost(x:Ref) : Bool

  axiom { forall r:Ref :: up_inv(up(r)) == r &&
    (is_ghost(r) ? up(r) == r : heap(r)==0 ==> heap(up(r)) == 1) }
  axiom { forall r:Ref :: {up_inv(r)} up(up_inv(r)) == r &&
    (is_ghost(r) ? up_inv(r) == r : heap(r)==1 ==> heap(up_inv(r)) == 0) }
  axiom { forall r:Ref :: {down(r)} down_inv(down(r)) == r  &&
    (is_ghost(r) ? down(r) == r : heap(r)==0 ==> heap(down(r)) == -1) }
  axiom { forall r:Ref :: {down_inv(r)} down(down_inv(r)) == r &&
    (is_ghost(r) ? down_inv(r) == r : heap(r)==-1 ==> heap(down_inv(r)) == 0) }

  axiom { forall r:Ref :: {temp(r)} temp_inv(temp(r)) == r  &&
    (is_ghost(r) ? temp(r) == r : heap(r)==0 ==> heap(temp(r)) == -3) }
  axiom { forall r:Ref :: {temp_inv(r)} temp(temp_inv(r)) == r &&
    (is_ghost(r) ? temp_inv(r) == r : heap(r)==-3 ==> heap(temp_inv(r)) == 0) }
}

$\encodeStm{\GhostAlloc{l}}\leadsto
$
  x := new(); assume ghostRef(x); // ghost location
  inhale $\Uninit{x}$ // ghost locations are non-atomics

\end{silver}
\vspace{-0.75em}
\caption{Extension of our Viper encoding to handle ghost locations.}
\label{fig:ghosts}
\end{figure}

We extend our encoding to handle ghost locations in a simple manner. Firstly, we add a boolean function \code{is_ghost} on references, to identify whether or not a location is ghost. We added macros \code{realRef(r)} to add the appropriate assumptions for a real program reference, and \code{ghostRef(r)} for ghost locations; this can be seen as the translation of ``type information'', since we assume that ghost locations are labelled as such in the original program.

For ghost locations we tweak our multiple heaps encoding to change the assumptions about the $\up$ and $\down$ mappings to instead require them to act as the identity (correspondingly, the result of $\heap$ is no longer constrained to be different after applying these mappings to a ghost reference). This immediately gives us that, for assertions depending only on ghost locations in the heap, $\UP{A}$, $A$ and $\DOWN{A}$ will be handled equivalently; since (up to syntactic applications of these identify mappings) they will be encoded as identical assertions.

Finally, we add an assumption of \code{realRef(r)} to our existing statements for allocating references, and add a new ghost allocation statement, for which the analogous \code{ghostRef(r)} assumption is added. The most-relevant details are summarised in \figref{ghosts}.

%

\section{Compare and Swap Details}
\label{sec:compare-and-swap-details}

\begin{figure}[t]
\begin{silver}[mathescape=true]
define SomeRMWAcq(l) acc(l.acq, wildcard) && l.acq == false

$\encodeAss{\RMWAcq{l,\Q}} \leadsto$ SomeRMWAcq(l) &&
  ($\foreach{i}{\conjuncts{\Q}}$: acc(AcqConjunct(l, $i$),wildcard) $\foreachEnd$)

$\encodeStm{\AtomicAllocRMW{l}{\Q}}\leadsto$
  x := new(); assume realRef(x); // not a ghost location
  inhale $\encodeAss{\Rel{l,\Q}}$ && $\encodeAss{\RMWAcq{l,\Q}}$

$\encodeStm{\CAS{x}{\tau}{l}{e}{e'}}\leadsto$
  assert Init(l) && SomeRMWAcq(l) && SomeRel(l)
  x := havoc()
  // inhale into tmp heap
  if(x == $\encodeAss{e}$) {
    $\foreachNumber$ $\foreachStart$
      if (perm(AcqConjunct(l, $i$)) > 0) {
        inhale $\instantiate{\tempAss{\invariant{i}}}{x}$
      }
    $\foreachEnd$
    // exhale from tmp and real/up heaps (depending on $\tau$)
    $\foreachNumber$ $\foreachStart$
      if ($i$ == l.rel) { // write synchronises
        if ($\tau\in\{\rel,\relacq\}$) {
          exhale $\instantiate{\tempOrRealAss{\invariant{i}}}{\encodeAss{e'}}$
        } else {
          exhale $\instantiate{\tempOrUpAss{\invariant{i}}}{\encodeAss{e'}}$
        }
      }
    $\foreachEnd$
    // ... move tmp heap to real/down heap (depending on $\tau$)
    var rs : Set[Ref]; ${}$ rs := havoc() // unknown set of Refs
    assume forall r: Ref :: r in rs <==> perm(tmp(r).val) > none
    if($\tau\in\{\acq,\relacq\}$) {
      inhale forall r: Ref :: r in rs ==> acc(r.val, perm(tmp(r).val))
      assume forall r: Ref :: r in rs ==> r.val == tmp(r).val
    } else {
      inhale forall r: Ref :: r in rs ==> acc(down(r).val, perm(tmp(r).val))
      assume forall r: Ref :: r in rs ==> down(r).val == tmp(r).val
    }
    exhale forall r: Ref :: r in rs ==> acc(tmp(r).val, perm(tmp(r).val))
  //$\textit{ analogously for each other field, predicate (in place of }$val$\textit{)}$
  }
\end{silver}
\vspace{-0.75em}
\caption{Viper encoding of the RSL rules for compare and swap operations.}
\label{fig:compare-and-swap-Viper}
\end{figure}
The details of our encoding of the \FSLpp{} compare and swap rules (cf.~\figref{compare-and-swap-RSL}) are shown in \figref{compare-and-swap-Viper}. We represent $\RMWAcqX$ assertions similarly to $\AcqX$ assertions (cf.~\figref{release-acquire-Viper}), but and a \texttt{false} value of the \texttt{acq} field to differentiate holding one from the other. Recall that we must choose at allocation whether atomic reads or compare and swaps will be used to gain ownership via the atomic location; this choice is then reflected in the field value. The encoding of allocation is then straightforward.

The handling of a CAS statement itself involves initially checking that we indeed hold some $\InitX$, $\RMWAcqX$ and $\Rel$ assertions for the location, according to the precondition of the rule, and then using an if-condition over the fresh read value \texttt{x} to narrow us down to the case of a successful CAS. The subsequent Viper code reflects the three steps described in \secref{compare-and-swap} and \figref{compare-and-swap-diagrams}. Firstly, we perform the inhale of newly-gained resources (corresponding to $\instantiate{\Q}{e}$) into the \texttt{\tempName} heap.

Secondly, we attempt to exhale the assertion $\instantiate{\Q}{e'}$, modified so that the permissions are taken preferentially from the \texttt{\tempName} heap, and failing this, from the real heap or \texttt{\upName} heaps, depending on whether or not the write synchronises. This modification of the assertion (which splits the permission amounts across the two heaps, as described in \secref{compare-and-swap}) is denoted by the $\tempOrRealAss{.}$ and $\tempOrUpAss{.}$ mappings; if the \emph{values} of heap locations are also mentioned in the parameter assertions, then these heap dereferences must also be rewritten to a dereference in the corresponding heap (e.g.~\texttt{x.val == 4} might become \texttt{\temp{x}.val == 4}). In case permission to the corresponding location is taken partly from both heaps, the extra assumption that the two values are the same can be explicitly added by these mappings.

Finally (assuming the exhale has succeeded, otherwise a verification failure will have been encountered), all remaining permissions in the \texttt{\tempName} heap are transferred to either the real or \texttt{\downName} heap, depending on whether the read synchronises.

\section{Soundness}
\label{sec:soundnessdetails}
We outline the soundness of our encoding via three key ingredients. Firstly (\secref{invariants}), we identify invariants on the \emph{states} of the Viper programs which are in the image of our encoding; these invariants hold before and after (but not necessarily during) the code-fragments generated by the encoding of a single source-level statement. The invariants encode fairly basic properties, such as the fact that the amounts of permission held to the \code{val} and \code{init} fields of a non-atomic location are always the same. We can show straightforwardly that these invariants are preserved by the Viper programs
generated by our encoding. Throughout our arguments, we make use implicitly of the fact (also assumed at the source level, and in the RSL logics themselves) that locations are known to be either non-atomic or atomic locations; this is indirectly reflected at the Viper level in terms of which permissions/predicates are held for the locations, but is only explicitly relevant for constructing the soundness argument itself.

Secondly (\secref{mapping}), for Viper states satisfying these invariants, we define a mapping from the state to an \emph{assertion} of the RSL logics. Conceptually, this mapping can be thought of as capturing where we are in the construction of a Hoare Logic proof in the original formalism. This is connected to our soundness argument by then showing that, if one compares the initial and final states of the encoding of any source-level statement, and applies our mapping to each, the assertions represent a Hoare triple derivable in the original logics \emph{provided that the Viper-encoded program has no verification errors}. Thus, we connect verification at the Viper level, with the construction of a proof at the Hoare logic level.

Finally (\secref{entailments}), we need to be sure that Viper does not, \eg{} deduce inconsistency at points in a proof where this would not be justified in the original logic. In general, we would like to know that any \emph{entailments} between assertions in a single state which Viper can justify automatically, reflect entailments which were justified in the original logic.

Putting these three ingredients together, we know that the verification of an encoded Viper program will imply the existence of a Hoare Logic derivation in the original logics; \ie{} that our encoding approach provides a sound mechanism for implementing the logics.

\subsection{Viper States and Semantics}
The states of a Viper program consist of a triple $(H,P,\env)$ of a \emph{heap} $H$ (mapping \texttt{Ref} and field name pairs to values), a \emph{permission map} $P$ (mapping such pairs, as well as \emph{predicate instances} to \emph{permission amounts}, which can be considered non-negative rational values; for field locations, these cannot exceed $1$), and an \emph{environment} $\env$, mapping variable names to values. We write $H[r,f]$ and $P[r,f]$ for lookups in these maps; for looking up \eg{} predicates $p(r)$ in the permission map, we write $P[p(r)]$.

 The semantics of the core logic is given in \cite{ParkinsonSummers12}; in particular, the semantics of heap-dependent expressions such as heap dereferences $x.f$ comes with a well-definedness condition; such heap dereferences are only allowed in states in which \emph{non-zero} permission is held (\ie{} $P[x,f] > 0$). The treatment of functions and predicates in the logic follows \cite{SummersDrossopoulou13}.

Verification of a Viper program amounts to two things: checking that all \texttt{assert} and \code{exhale} statements describe assertions valid in the corresponding state (both are sources of verification failures; the difference is that any permissions/predicates in the parameter to an \texttt{exhale} statement are also removed by the end of the statement), and checking that all expressions employed in the program are \emph{well-defined}: for heap dereferences, this means checking that some permission to the corresponding location is held, while for application of specification \emph{functions} (such as \code{valsRead} in our encoding), this means checking that their preconditions hold where they are applied. Some assertions are implicitly defined via specifications: for example, a method postcondition must be shown to hold at the end of the method body.

\newcommand{\val}{\texttt{val}}
\newcommand{\init}{\texttt{init}}
\newcommand{\falseval}{\textit{false}}
\newcommand{\trueexp}{\textit{true}}
\newcommand{\AcqConjunct}{\texttt{AcqConjunct}}
\subsection{Invariants}\label{sec:invariants}
Apart from the classification of references into those representing non-atomic and atomic locations, our argument depends on the following invariants on states $(H,P,\env)$, guaranteed to hold at the start and end of each block of Viper code representing the encoding of a single source-level statement:
\[\begin{array}{l}
\textit{For non-atomic locations }l:\\
P[l,\val] = P[l,\init] \wedge (P[l,\val] > 0 \wedge H[l,\init] = \falseval \Rightarrow P[l,\val] = 1)\\
\end{array}
\]
It is straightforward to show that these invariants are preserved by our statement encoding cases; for example, allocation of a non-atomic location provides full permission to both $\val$ and $\init$ fields; these permissions can only be given away by $\Init{l}$ and points-to assertions, whose encodings (see \secref{source-encoding}) also require identical permission amounts to both fields.

\subsection{Mapping and Hoare Triples}\label{sec:mapping}
We next define the mapping $\concrete{l}{H}{P}{\env}$ from a reference $l$ in a Viper state $(H,P,\env)$ (which is assumed to satisfy the invariants in \secref{invariants}) to \emph{assertions} from the RSL logics; the corresponding mapping for the entire Viper state is then the iterated separating conjunction \cite{Reynolds02a} over the assertion for each reference to which at least some permission is held.

 We deal concretely with the simplified case of the logics without the $\UP$ and $\DOWN$ modalities, and then explain how to extend the definitions.

For non-atomic locations $l$, the mapping is defined as follows:
\[
\concrete{l}{H}{P}{\env} = \left\{\begin{array}{ll}
\Uninit{l} &\textit{ if }H[l,\init] = \falseval\\
\pointsto{l}{k}{v} &\textit{ otherwise, where }v = H[l,\val]\textit{ and }k = P[l,\val]
\end{array}\right.
\]
We allow ourselves here the technical liberty of ``re-inserting'' an integer value $v$ as a logical variable in the resulting assertion.

\makeatletter
\DeclareRobustCommand\bigop[1]{%
  \mathop{\vphantom{\sum}\mathpalette\bigop@{#1}}\slimits@
}
\newcommand{\bigop@}[2]{%
  \vcenter{%
    \sbox\z@{$#1\sum$}%
    \hbox{\resizebox{\ifx#1\displaystyle.9\fi\dimexpr\ht\z@+\dp\z@}{!}{$\m@th#2$}}%
  }%
}
\makeatother
\newcommand{\itsep}[1]{\ensuremath{\bigop{\ast}_{#1}}}
\newcommand\vecmap[3]{\ensuremath{#1\vec{[#2\mapsto#3]}}}
For non-atomic locations $l$, the mapping is more involved:
\[
\begin{array}{rcl}
\concrete{l}{H}{P}{\env} &=& \cond{P[l,\init] = 0}{\trueexp}{\Init{l}} \ast\\
&& \cond{P[l,\rel] = 0}{\trueexp}{\Rel{\inv{H[l,\rel]}}} \ast \\
&& (P[l,\acq]=0~{?}~\trueexp~{:}~(H[l,\acq]=\trueexp~{?}\\
&&\quad\Acq{\itsep{i\mid P[\AcqConjunct(l,i)] \geq 1}\vecupd{\inv(i)}{j}{j\in\eval{\texttt{valsRead}(l,i)}{H}{P}{\env}}}~{:}\\
&&\quad\RMWAcq{\itsep{i\mid P[\AcqConjunct(l,i)] \geq 1}\inv(i)}))
\end{array}
\]
Here, we rewrite $\eval{\texttt{valsRead}(l,i)}{H}{P}{\env}$ for the semantics of this function application in the given state; \ie{} the set of integer values it represents.

In brief, the above mapping reconstructs an appropriate $\Init$, $\Rel$, and either $\Acq$ or $\RMWAcq$ assertion for the corresponding location, according to the permissions (and predicates) held in the state.

The mappings above can be generalised to the full logics with modalities by reflecting on the $\heap$ numbering of the reference in question (\cf{} \secref{fences}); where $\heap{l} = 0$, the above definitions apply, while for $1$ or $-1$ the resulting assertion must be placed under the $\UP$ or $\DOWN$ modalities, respectively.

For each source language statement, one can now show that \emph{if} the encoded Viper statements verify, the beginning and end states of the Viper program must, when the above mapping is applied, describe a provable Hoare triple in the original logic.

\subsection{Entailment Correspondence}\label{sec:entailments}
In addition to the encoding of individual statements, it is important to consider which entailments Viper can automatically prove about the encoded assertions from the original logics. For the assertions describing non-atomic locations, Viper's built-in field permissions are used in a standard manner; the relationship between the handling of these permissions in such a logic and a typical concurrent separation logic presentation is well-understood to give an isomorphism \cite{ParkinsonSummers12}. In particular, Viper imposes the same assumptions for field permissions (that no more than $1$ permission can be held) as in a standard separation logic.

 For the encoding of non-atomic locations, the Viper representation is largely in terms of duplicable (wildcard) permissions, and abstract predicates. Wildcard permissions, as discussed in \secref{release-acquire}, model a duplicable resource exactly as desired. Abstract predicates, on the other hand, are treated as unknown resources in Viper; these are counted in and out when inhaled and exhaled, but no additional facts will be deduced from holding them in a particular state. Our modelling of atomic invariants with $\AcqConjunct$ predicates can, in some cases, provide entailments between the encodings of different $\Acq$ predicates, but these are always instances of the general rules of the logic.
\end{appendix}

\end{document}